\documentclass[a4paper,11pt,headings=big,DIV=12]{scrartcl}
\usepackage{amsmath,amssymb}
\usepackage{color,xcolor}
\definecolor{blue}{rgb}{0,0,0.5} 
\usepackage[colorlinks,linkcolor=blue,urlcolor=blue,citecolor=blue]{hyperref}
\usepackage{graphicx}
\addtokomafont{disposition}{\rmfamily\boldmath}
\usepackage[utf8]{inputenc}
\usepackage{slashed}
\usepackage{multirow}
 \usepackage{bbold}
 \usepackage{mathbbol}

\pdfoutput=1 

\usepackage{bbm}

\usepackage[utf8]{inputenc}
\allowdisplaybreaks
\DeclareMathAlphabet{\mathantt}{OT1}{antt}{li}{it}
\DeclareMathAlphabet{\mathpzc}{OT1}{pzc}{m}{it} 

\usepackage{simplewick} 

\usepackage{makecell}	
\usepackage{tabu}		

\usepackage[makeroom]{cancel} 

\usepackage[normalem]{ulem} 

\usepackage{amsthm}

\usepackage[utf8]{inputenc}

\usepackage{amsmath,amsfonts,amsthm,amssymb} 

\numberwithin{equation}{section}

\newcommand{\p}{\partial}

\newcommand{\eq}{&\quad}

\newcommand{\rig}{\right.}
\newcommand{\lef}{\left.}

\newcommand{\mco}{\mathcal{O}}

\newcommand{\para}{\parallel}
\newcommand{\hD}{{\hat{\Delta}}}
\newcommand{\hp}{{\hat{\phi}}}
\newcommand{\hg}{\hat{g}}

\newcommand{\vev}[1]{\langle #1 \rangle}

\newcommand{\al}{\alpha}
\newcommand{\be}{\beta}

\newcommand{\de}{\delta}
\newcommand{\e}{\epsilon}
\newcommand{\ph}{\phi}
\newcommand{\g}{\gamma}

\newcommand{\m}{\mu}

\newcommand{\oo}{\omega}

\newcommand{\x}{\xi}

\newcommand{\D}{\Delta}

\newcommand{\G}{\Gamma}
\newcommand{\La}{\Lambda}

\begin{document}

 \thispagestyle{empty}

\begin{flushright}
\begin{tabular}{l}
UUITP-51/19\\
\end{tabular}
\end{flushright}
\vskip1.5cm

\begin{center}
{\Large\bfseries \boldmath Composite operators near the boundary}\\[0.8 cm]
{\Large%
Vladimír Procházka, Alexander Söderberg
\\[0.5 cm]
\small
 Department of Physics and Astronomy, Uppsala University,\\
Box 516, SE-75120, Uppsala, Sweden 
} \\[0.5 cm]
\small
E-Mail:
\texttt{\href{mailto:vladimir.prochazka@physics.uu.se}{vladimir.prochazka@physics.uu.se}},
\texttt{\href{mailto:alexander.soderberg@physics.uu.se}{alexander.soderberg@physics.uu.se}}.
\end{center}

\bigskip

\pagestyle{empty}
\begin{abstract}

We use renormalization group methods to study composite operators existing at a boundary of an interacting  conformal field theory. In particular we relate the data on boundary operators to short-distance (near-boundary) divergences of bulk two-point functions. We further argue that in the presence of running couplings at the boundary the anomalous dimensions of certain composite operators can be computed from the relevant beta functions and remark on the implications for the boundary (pseudo) stress-energy tensor.  We apply the formalism to a scalar field theory in $d=3-\epsilon$ dimensions with a quartic coupling at the boundary whose beta function we determine to the first non-trivial order. We study the operators in this theory and compute their conformal data using $\epsilon-$expansion at the Wilson-Fisher fixed point of the boundary renormalization group flow. We find that the model possesses a non-zero boundary stress-energy tensor and displacement operator both with vanishing anomalous dimensions. The boundary stress tensor decouples at the fixed point in accordance with Cardy's condition for conformal invariance. We end the main part of the paper by discussing the possible physical significance of this fixed point for various values of $\epsilon$.

\end{abstract}

\newpage 

\setcounter{tocdepth}{3}
\setcounter{page}{1}
\tableofcontents
\pagestyle{plain}

\newtheorem{defin}{Definition}
\newtheorem{thm}{Theorem}
\newtheorem{cor}{Corollary}
\newtheorem{pf}{Proof}
\newtheorem{nt}{Note}
\newtheorem{ex}{Example}
\newtheorem{ans}{Ansatz}
\newtheorem{que}{Question}
\newtheorem{ax}{Axiom}

\section{Introduction}

\subsection{From operator product expansion to renormalization}

Composite operators (i.e local polynomials of fundamental fields at coincident points) play an important role in understanding the properties of strongly coupled systems. In particular they can be used to parametrize bound states that are not accessible via conventional perturbative expansions. Of particular interest are certain special operators such as conserved Noether currents, which are independent of the renormalization group (RG) flow and therefore provide and important window into the observable physics.

\quad\quad The difficulty in defining the composite operators in a generic quantum field theory (QFT) with a given Lagrangian stems from the short-distance divergences that appear when the two fundamental fields come close to each other \cite{Coleman:1969zi}. A singularity like this usually cannot be renormalized by field redefinition and has to be understood by other methods such as operator product expansion (OPE).

\quad\quad The intimate connection between renormalization of short-distance divergences and OPE goes back to the seminal work of Wilson \cite{Wilson:1965zz} and Wilson and Zimmerman \cite{Wilson:1972ee}. In these works the authors used OPE of fundamental fields $\phi$ as a tool for defining the composite operators
\begin{equation} \label{eq:WilsonOPE}
\phi(x-\xi)\phi(x+\xi) \stackrel{\xi \to 0}{\sim}  C_{\phi^2}(\xi) [\phi^2(x)] + \dots \; ,
\end{equation}
where $C_{\phi^2}(\xi)$ (also called Wilson coefficient) typically diverges logarithmically in the $|\xi| \to 0$ limit and dots include power divergent contributions of other operators of lower engineering dimension (which includes the unity operator as well).\footnote{In general there can also be a mixing between different operators with the same engineering dimension.} For a conformal field theory (CFT) the OPE coefficients enjoy a universal power law behaviour dictated by the scale invariance \cite{Wilson:1970pq}, whereas in a generic QFT they depend on the short distance physics (i.e running couplings, anomalous dimensions near the UV fixed point of the theory).  The square brackets around $[\phi^2]$ indicate that it is a renormalized operator in the sense that its single point insertion 
\begin{equation} \label{eq:RenormFinite}
\vev{[\phi^2] \dots}  \; 
\end{equation}
into renormalized correlator $\vev{\dots}$ remains finite. The l.h.s. of \eqref{eq:WilsonOPE} can therefore be understood as a bare composite operator with $C_{\phi^2}(\xi)$ encoding the short distance (UV) divergences. Indeed this approach is commonly used to define composite operators in lattice simulations where $\xi$ has a natural interpretation of the lattice spacing \cite{Martinelli:1987bh}.
In models that are weakly coupled at short distances, one can compute the Wilson coefficients perturbatively\footnote{For example this observation was successfully applied in determination of sum rules in the  quantum chromodynamics (QCD) \cite{Shifman:1978bx}.} and therefore determine the renormalized composite operator as
\begin{equation} \label{eq:BulkWilsonRenorm}
[\phi^2(x)] \equiv \lim_{\xi \to 0} \frac{ \phi(x-\xi)\phi(x+\xi)- \dots}{C_{\phi^2}(\xi)} \; ,
\end{equation}
where the dots indicate that the power divergences have been subtracted. 
The scale dependence of $[\phi^2(x)]$ can be deduced using a following standard Wilsonian argument. In a perturbative regime
\begin{equation}
C_{\phi^2} \sim 1 + c g(\mu) \log(\mu \xi) + \mathcal{O}(g^2) \; ,
\end{equation}
where $\mu$ is the IR cutoff scale specific to the problem at hand that has to be introduced to keep the arguments of logarithms scale-free and $g$ is some (small) coupling. The l.h.s. of \eqref{eq:BulkWilsonRenorm} depends only on $\mu$, whereas the dependence on the UV cutoff $\xi$ drops out as expected.
The anomalous dimension of $[\phi^2(x)]$ is then encoded in the $\mu-$dependence of $C_{\phi^2}$.\footnote{More precisely $C_{\phi^2}$ determines $\gamma_{\phi^2}- 2 \gamma_\phi$ as we still have to take into account the 'naive' dimension of the product of $\phi$s at non-coincident points in the numerator of \eqref{eq:BulkWilsonRenorm}.} Hence we see that the RG controls the UV behaviour of operator products. This connection can be seen by examining the large momentum structure of composite operator correlators \cite{Prochazka:2016ati}. We will argue in the present paper that similarly the near-boundary behaviour of correlators is controlled by the RG group.

\quad\quad By contrast if we choose to regulate the UV divergences of \eqref{eq:BulkWilsonRenorm} via dimensional regularization (dimreg), the power-divergent terms vanish by construction and we are left with a dimensionless coefficient $C_{\phi^2} = \frac{1}{\epsilon} + finite$. This divergence can then be dealt with by renormalization of the source corresponding to $[\phi^2]$. The composite operator we obtain this way is not the same as \eqref{eq:BulkWilsonRenorm}, however the physical data such as its anomalous dimension at the fixed point have to agree. Note that in the cases where the operator in question is protected (such as stress-energy (SE) tensor or conserved currents) the limit \eqref{eq:BulkWilsonRenorm} is finite giving a unique definition of renormalized product up to a multiplicative numerical constant.
 
\quad\quad Let us now turn the attention to composite operators in a theory with boundary. The motivation to study such theories in field theory comes mostly from condensed matter physics where they describe the surface critical phenomena (we refer the reader to \cite{Diehl:1996kd} for an excellent overview).  Here we will consider a $d-$dimensional CFT with a (planar) boundary  at $x_\perp=0$ in some coordinate system $x \equiv (x_\parallel^a, x_\perp)$, where indices $a$ run from $1$ to $d-1$ and $x_\perp \geq 0$. It is well known that conformal symmetry implies UV-finiteness, however the presence of boundary breaks the conformal group $SO(d+1,1)$ to a subgroup, which might not even include the scale transformations (at best we can get the boundary conformal group $SO(d,1)$) .
An insertion of a \textit{bulk} operator $[O(x)]$ in \eqref{eq:RenormFinite} will no longer remain finite in the \textit{near-boundary} limit $x_\perp \to 0$. In this regime one usually employs the boundary operator expansion (BOE) \cite{Diehl1981,CARDY1991274, 9505127}
\begin{equation} \label{eq:BOEintro}
[O(x)]= \sum_{k} \mu_k(x_\perp) \hat{O}_k(x_\parallel) \; ,
\end{equation}

where $\hat{O}_k$ are renormalized boundary operators, whose single point insertions \eqref{eq:RenormFinite} are finite.\footnote{The appearance of boundary divergences can also be understood from the OPE perspective using the following qualitative picture. From the method of images we expect the operator product \eqref{eq:WilsonOPE} to have a contribution of the form
\begin{equation*} 
\phi(x-\xi)\phi(\bar{x}+\bar{\xi}) \; ,
\end{equation*}
where $\bar{x}= (x_\parallel, - x_\perp)$ is the position of the image. If we now take the coincident limit $\xi \to 0$ this term remains finite as long as $x_\perp>0$ and can be expanded using OPE for small $x_\perp = \frac{1}{2}|\bar{x}-x|$
\begin{equation*}
\phi(x)\phi(\bar{x})\sim \sum_k \tilde{C}_k(x_\perp) O_k (x_\parallel, x_\perp = 0) \; ,
\end{equation*}
which is exactly of the form \eqref{eq:BOEintro}.} In a b(oundary)CFT (we will explain later on what we mean by that) $\mu_k(x_\perp)$ behave as $x_\perp^{\Delta_{\hat{O}}-\Delta_{O}}$, where $\Delta$s denote the dimensions of respective operators. The coefficients $\mu_k(x_\perp)$ generically diverge in the near boundary limit \cite{Diehl:1996kd} and these divergences can be renormalized in dimreg by introducing a new set of counterterms localised at the boundary \cite{Diehl1981, Symanzik:1981wd}. The latter two works introduced the RG formalism to boundary theories as a tool to understand their physical properties independently of the underlying symmetries. On the other hand a lot of progress has been achieved by exploiting the boundary conformal symmetry to restrict the form of bulk two-point functions \cite{9505127, 9302068}. These ideas were further developed  using the conformal bootstrap wherewith one can relate the BOE coefficients to the bulk conformal data \cite{Liendo:2012hy}. While these techniques are very powerful they still require the use of conformal symmetry, which might not be always available (for example if we introduce boundary conditions which break scale invariance). Furthermore it is not always clear how to properly define boundary conditions for composite operators given the r.h.s. of \eqref{eq:BOEintro} is not necessarily well defined. 
In this paper we will use \eqref{eq:BOEintro} as a tool to define composite operators at the boundary from the bulk ones in an analogy with \eqref{eq:BulkWilsonRenorm}. While this approach might not be new in itself our aim is to create a bridge between more traditional field theory approach of \cite{Diehl1981, Symanzik:1981wd} and modern conformal bootstrap techniques \cite{1808.08155, Kaviraj:2018tfd, Mazac:2018biw}. We will clarify how one can compute the data of boundary operators from divergent bulk correlators. In our setup we will assume existence of running relevant/marginal couplings at the boundary, which break the conformal symmetry everywhere but at the fixed points of the RG flow. A crucial and novel part of this paper is the interacting model we introduce in Section \ref{sec:BoundaryExample}. This model is free in the bulk and has a single relevant coupling at boundary so it shares some essential properties with the models for graphene that were recently studied in the literature (see \cite{1807.01700, DiPietro:2019hqe, Grignani:2019zxc}). The physical significance of the model introduced in this paper is more in relation to the boundary Landau-Ginzurg models \cite{Cappelli:2003ct}, which are relevant for condensed matter physics \cite{Giuliano:2005tbr} as well as string theory \cite{Kutasov:2000qp}.
To our best knowledge this model (or by very least the analysis of composite operators in it) has not appeared in the literature. We will show that it has a few interesting properties such as a beta function \eqref{eq:FinalBeta} admitting an interacting fixed point and non-renormalization of the boundary fields \eqref{eq:phiNonRen}. There will be composite operators that acquire anomalous dimensions through renormalization of near-boundary divergences. In this theory there is also a non-trivial boundary SE tensor of the form \eqref{eq:tabRen} which is analogous to the usual trace anomaly of bulk QFT. 

\subsection{Outline}

In Section \ref{sec:bCFT} we will introduce the notation and formalism of the paper. In particular we will discuss the kinds of divergences that arise in two-point correlators of bulk operators and where they come from in section \ref{sec:Divergences}. Here we also clarify 
how to define finite boundary operators by renormalizing the boundary limit of bulk operators and argue that the boundary conditions actually relate bare/divergent operators and how this can be related to renormalization of boundary couplings. In Section \ref{sec:Dimreg} we give a recipe how to properly introduce dimensional regularization in presence of a boundary and use it to derive RG equations. In Section \ref{sec:CutoffReg} we relate \eqref{eq:BOEintro} to the near-boundary divergences of two-point functions and use this framework in Section \ref{sec:BulkModel} to identify composite operator contributions to a correlator found by conformal bootstrap. In Section \ref{sec:SEtensorGeneral} we describe the emergence of boundary SE tensor in the presence of classically marginal scalar operators at the boundary. 

\quad\quad In Section \ref{sec:BoundaryExample} we present the model we will work with in this paper. Scalars in the bulk of this three dimensional model are free, while on the boundary there is a quartic interaction. The running of its coupling constant, namely its beta function and fixed points  are studied in Section \ref{sec:BetaDeriv}. In Section \ref{sec:PhiSquared} we illustrate how one can apply dimreg as well as an infinitesimal boundary cutoff to deal with the near-boundary divergences of $\phi^2$. In Section \ref{sec:PhiFour} we perform two consistency checks of the general formalism we introduced in Section \ref{sec:bCFT} and study the renormalization of $\phi^4$.

\quad\quad We proceed to study the SE tensor for the model in Section \ref{Sec:SETensor}. We first describe how it is found from variational principles  then determine the boundary pseudo-SE tensor from RG in Section \ref{sec:TenRen} and verify that the displacement operator (which is the normal-normal component of the SE tensor) is indeed a protected operator in Section \ref{sec:Disp}.  In Section \ref{sec:SE} we perform a consistency check on the formulas derived in Section \ref{sec:TenRen}.

\quad\quad Lastly, we will discuss the potential physical relevance of the fixed point we found and outline in which direction one can apply and generalize the methods explored in this paper in Section \ref{sec:Outlook}.

\section{CFT with a boundary} \label{sec:bCFT}

\subsection{Setup of the paper}

We will now consider the same setup as around \eqref{eq:BOEintro} with the bulk being $d-$dimensional CFT having a planar boundary at $x_\perp=0$. The boundary theory will be defined by a set of boundary conditions imposed on the bulk fields. In a generic case the theory will not have the full boundary conformal symmetry $SO(d, 1)$\footnote{This group includes scale transformations, $d-1$ special conformal transformations, $d-1$ translations and the group of rotations along the boundary $SO(d-1)$.} but rather just the subgroup including the parallel translations and rotations. This can happen in presence of boundary conditions breaking scale/conformal invariance. The bulk theory will be defined by a set of local operators $\{O_i\}$ with scaling dimensions $\{\Delta_i\}$ subject to boundary conditions (we will describe how to impose them in the next section). The bulk couplings are assumed to be tuned to a fixed point or belong to a conformal manifold. On top of that we will also introduce a set of boundary couplings $\{\hat{g}_I \}$, which couple to operators $O_I(x_\perp=0)$ that are assumed to be relevant/marginal at the boundary (i.e. $\Delta_I \leq d-1$). This theory has a conserved, traceless bulk SE tensor $T_{\mu \nu}$ satisfying
\begin{equation}
\partial_\mu T^{\mu \nu} = 0 \ , \quad T^\m{}_{\m} = 0 \ .
\end{equation}

In addition the invariance w.r.t. parallel translations imposes a boundary condition
\begin{equation} \label{eq:MomentumCond}
\lim_{x_\perp \to 0} T^{a \perp}(x) = \partial_b \hat{\tau}^{ ab}(x_\parallel) \; ,
\end{equation}

where $\hat{\tau}^{ ab}$ is a symmetric tensor operator of (engineering) dimension $(d-1)$ defined at the boundary. Alternatively $\hat{\tau}^{ ab}$ can be introduced  by coupling it to the boundary metric \\
$\hat{g}_{ ab} \equiv g_{ab}|_{x_\perp = 0}$ and performing a variation of the curved space action $S$
\begin{equation}
\hat{\tau}^{ ab}(x_\para) = -\frac{2}{\sqrt{\hat{g}}}\frac{\delta S}{\delta \hat{g}_{ab}} \Big{|}_{\hat{g}_{ ab}= \delta^{ab}} \; .
\end{equation} 
The full stress-energy tensor generating spacetime symmetries of the bulk \textit{and} the boundary is then defined by adding a distributional contribution $\delta(x_\perp) \hat{\tau}^{ ab} $ to the parallel components of the SE tensor. 

\quad\quad Another component of interest is
\begin{equation}\label{eq:DisplacementDef}
\lim_{x_\perp \to 0} T^{\perp \perp}= \hat{D}(x_\parallel) \; ,
\end{equation}

where on the r.h.s. we have the displacement operator, which describes the infinitesimal deformations of the boundary in the normal direction. More concretely under diffeomorphisms $\delta x_\perp = \zeta(x_\para)$ perpendicular to the boundary one defines 
\begin{equation} \label{eq:DisplacementDef1}
\hat{D}(x_\para)= \lef \frac{\delta S}{\delta \zeta(x_\para)} \right|_{\zeta=0}  \; .
\end{equation}

If we want a theory to have the boundary conformal symmetry the SE tensor has to satisfy a stronger condition \cite{CARDY1984514}
\begin{equation} \label{eq:CardyCon}
\lim_{x_\perp \to 0} T^{a \perp} =\partial_b \hat{\tau}^{a b}(x_\parallel)= 0 .
\end{equation}

Theories satisfying \eqref{eq:CardyCon} are what one usually denotes as a bCFT. For such theories the BOE \eqref{eq:BOEintro} takes the form
\begin{equation} \label{eq:BOEcft}
O(x) \sim \sum_{k} \mu_k x_\perp^{( \hat{\Delta}_{O_k}-\Delta_O)} \hat{O}_k(x_\parallel) \; 
\end{equation}

where the $\hat{O}_k$s in this sum are primaries. The contribution of descendants can be incorporated by including derivative-dependent BOE coefficients just as one would do in a usual OPE (see Appendix \ref{app:BulkBdyCorr} for more details).  The residual conformal symmetry together with BOE can be used to fix the form of correlators \cite{9505127}. 
A boundary-boundary correlator behaves as a two-point function in a homegenous CFT 
\begin{equation} \label{eq:TwoPtNorm}
\begin{aligned}
\langle\hat{O}_i(x)\hat{O}_j(y)\rangle &= \frac{\de_{ij}}{\left|s_\para\right|^{2\hD}} \ ,
\end{aligned}
\end{equation}
where we $\textit{chose}$ a particular normalization of $\hat{O}$s. 
 Bulk-boundary correlators are similar to a three-point function in a homegenous CFT. Their follows from the residual $SO(d - 1)$-symmetry preserved by the boundary, and it can be found from the BOE\footnote{See Appendix \ref{app:BulkBdyCorr} for more details on this.}
\begin{equation} \label{eq:BoundaryCorrelator}
\begin{aligned}
\langle O(x)\hat{O}(y)\rangle = \frac{\m^O{}_{\hat{O}}}{\left|x_\perp\right|^{\D - \hD} \left( s_\para^2 + x_\perp^2 \right)^{\hD}} \ .
\end{aligned}
\end{equation}

\subsection{Two-point functions and divergences} \label{sec:Divergences}
For clarity we will look at a two-point function of two bulk operators without a spin
\begin{equation} \label{eq:BulkTptFn}
\vev{O_i(x) O_j(y)} \; .
\end{equation}

This correlator will involve two kinds of divergences:
\begin{itemize}
\item{\textit{Bulk UV divergences}}\\
These are corresponding to the short-distance (large momentum) infinities arising from internal loops in \eqref{eq:BulkTptFn} for $x_\perp, y_\perp >0$.\footnote{In general one can also encounter contact term divergences that appear in the presence of local sources. Here we will assume that $x \neq y$ so such divergences will not play a role in the discussion. For more discussion of how to deal with contact terms in a bCFT we refer the reader to \cite{1804.01974}.}
 They can be can be dealt with by renormalizing the boundary couplings and by performing the usual bulk renormalization
 \begin{equation} \label{eq:BulkOren}
 Z_{O_i} O_i(x)= [O_i(x)] \; ,
\end{equation}  
for a finite, renormalized \textit{bulk} operator $[O(x)]$. Since the bulk is a CFT the factor $Z_{O}$ is simply proportional to $\mu^{-\gamma_{O}}$, where $\gamma_{O}$ is the anomalous dimension of $O$.
The new feature of boundary theories is that in a non-conformal theory also the boundary couplings have to be renormalized
\begin{equation}
\hat{g}_I \to Z_{g_I} \hat{g}_I \; .
\end{equation}
In a bCFT satisfying \eqref{eq:CardyCon} the divergences related to coupling renormalization cancel out.
\item{\textit{Near-boundary divergences}}\\
Then we will generically have near-boundary divergences mentioned in the introduction, which will appear if we take the boundary limit of one or more points in the bulk correlator. More specifically the divergence appears when the following limit is taken
\begin{equation} \label{eq:Bulk2pGeneric}
\lim_{y_\perp \to 0}\vev{[O_i](x)[O_j](y)} \to \infty  \; ,
\end{equation}
where $[O_i]$ is renormalized as in \eqref{eq:BulkOren} and all the couplings (including boundary ones) have been renormalized as well. This type of singularity is therefore unrelated to the bulk UV divergences outlined above and has to be treated separately by introducing extra renormalization constants on top of \eqref{eq:BulkOren}.
\end{itemize}
The main assumption of this paper motivated by the work \cite{Diehl:1981jgg} is that one can think of bulk operators $O_i$ in the boundary limit as one would of bare operators in a usual QFT. One then defines the renormalized, finite boundary operators through 
\begin{equation} \label{eq:RenormFormula}
\lim_{y_\perp \to 0} Z_i{}^j \vev{O(x)O_j(y)} \equiv \vev{O(x) \hat{O}_i(y_\parallel)} = finite \; ,
\end{equation}

where $Z_i{}^j$ are typically divergent constants and $ \hat{O}_i$ are assumed to be renormalized boundary operators. 
This assumption is motivated by \eqref{eq:BOEintro} from which we expect that each $\hat{O}$ can be obtained from a BOE of a bulk operator (or its normal derivatives).\footnote{See footnote 2 of the published version of \cite{Lauria:2018klo}.}
The boundary couplings are introduced to the bulk Lagrangian $\mathcal{L}$ via
\begin{equation} \label{eq:BoundCouplingDef}
\mathcal{L} \to \mathcal{L}+ \delta(x_\perp) \hat{g}_0^I O_I(x_\perp=0)  \; ,
\end{equation}

where $\hat{g}_0$ and $O_I(x_\perp=0) \equiv \lim_{x_\perp \to 0} O_I(x)$ are to be thought of as bare (divergent) quantities. In particular this means that in an interacting theory boundary conditions should be interpreted as relations between bare operators.\footnote{This should be contrasted with equations of motion and Ward identities in a usual QFT, were they simply relate different divergent quantities.}
Following \cite{Brown:1979pq} we can obtain the renormalization constants for the operators $\{O_I \}$ by differentiating the path integral w.r.t. renormalized couplings $\{\hat{g}_I \}$.\footnote{In general one has to promote couplings to space-dependent background fields, which then act as sources for renormalized composite operators. Terms involving derivatives of sources that lead to mixing with descendant operators have to be included as well \cite{Osborn:1991gm}.} For operators coupling through \eqref{eq:BoundCouplingDef} this means that
\begin{equation} \label{eq:CouplingCond}
Z_I{}^J= \frac{\partial}{\partial \hat{g}^I}\hat{g}_0^J \; ,
\end{equation}

where $Z_I{}^J$ and the respective renormalized boundary operators are defined as in \eqref{eq:RenormFormula}.

\quad\quad We can regulate these divergences either by introducing a cutoff or by dimreg. There are pros and cons to using either of these regulators. 
The cutoff method has the advantage of connection to the BOE expansion \eqref{eq:BOEintro}. To regulate both types of divergences mentioned above we actually need two independent cutoffs in principle. One cutoff to regulate the UV divergences of boundary couplings and the other one regulating near-boundary divergences. The former can be avoided if we tune the couplings to a fixed point. 
Instead the dimreg regulates both types of divergences and does not involve power divergences. Of course the actual physical data, such as anomalous dimensions of boundary operators will not depend on the choice of regulator. In the next section we will discuss how to use each of these regulators to extract the conformal data in practical computations.

\subsection{Dimensional regularization and renormalization group equations} \label{sec:Dimreg}
Here we will only consider the two point function involving one bulk operator $O$ with scaling dimension $\Delta_O$ for illustrative purposes. It was argued (and verified perturbatively up to two loops) in \cite{Diehl:1981jgg} that one can regulate the both types of short distance divergences mentioned in the previous section using dimreg. The further advantage of dimreg is the absence of power divergences that would arise from closing the propagators on themselves.\footnote{\label{Footnote:PowerDiv} In some more detail the problem arises from the image part of the free propagator since
\begin{equation*}
\lim_{x \to y} \Delta_{image}(x,y) \sim \frac{1}{x_\perp^{(d-2)}} \stackrel{x_\perp \to 0}{\to} \infty \; .
\end{equation*}
Instead if we take the boundary limit first and only then evaluate the coincident limit we get
\begin{equation*}
\lim_{x_\perp, y_\perp \to 0} \Delta(x,y) \sim \frac{1}{|x_\parallel-y_\parallel|^{(d-2)}} 
\stackrel{x_\parallel \to y_\parallel}{\to} \int d^{d-1}k_{\parallel} \frac{1}{|k_\parallel|} \stackrel{dimreg}{=}0 \; .
\end{equation*}}

\quad\quad The bare operator is defined by taking $y_\perp \to 0$ limit of \eqref{eq:BulkTptFn}, which will include $\epsilon$ poles
\begin{equation} \label{eq:DimregOOcorr}
\vev{O(x) O(y_\parallel,0)}|_{dimreg}= \frac{1}{\epsilon}f(s_\para^2, x_\perp) + \dots \; ,
\end{equation}
where $f$ is a finite function of the parallel distance $s_\para^a \equiv x_\para^a - y_\para^a$ and $x_\perp$.

\paragraph{Important remark}{When regulating the correlators via dimreg the order of limits matters. In defining \eqref{eq:DimregOOcorr} we took the boundary limit ($y_\perp \to 0$) \textit{first} and only then epsilon-expanded. If we instead started with a bulk expression and expanded it in $\epsilon$ first, the boundary limit would not be well defined even for nonzero $\epsilon$. In the latter case we have to introduce a boundary cutoff, which is what we will do in the next section. When evaluating the correlators involving normal derivatives, we always take the derivative first and then the boundary limit and only then expand in $\epsilon$. }\\

The pole in \eqref{eq:DimregOOcorr} includes both types of divergences described in the previous section. It can be absorbed \cite{Diehl1981} into two \textit{independent} $Z-$factors and \eqref{eq:RenormFormula} becomes
\begin{equation} \label{eq:BulkEpsLimit}
 (Z_{O} \hat{Z}_{O}{}^{\hat{O}})O(y_\parallel,0) \equiv \hat{O}(y_\parallel) \;,
\end{equation}

where $Z_{O}$ is the bulk constant defined in \eqref{eq:BulkOren} and $\hat{O}$ is a boundary operator of the same engineering dimension as $O$ (for simplicity we assume there is only one such boundary operator).   In the BOE language the divergences in \eqref{eq:BulkEpsLimit} correspond to the logarithmically divergent contributions to \eqref{eq:BOEintro}. The factor $\hat{Z}_{O}$ is a dimensionless constant encoding the near-boundary divergences  that depends on the RG scale $\mu$ through the running of the couplings. I.e. the corresponding RG equation reads
\begin{equation} \label{eq:AnDimDimreg}
-\gamma_{\hat{O}}=  \frac{d}{d \ln \mu} \ln{\left(Z_{O} \hat{Z}_{O}{}^{\hat{O}}\right)}= -\gamma_{O}+\sum_{g^I, \hat{g}^I} \beta_{g^I} \frac{\partial}{\partial g^I} \ln{\left(\hat{Z}_{O}{}^{\hat{O}}\right)} \; ,
\end{equation} 
where the sum runs over both bulk and boundary couplings.\footnote{The conformal fixed point limit $g^I \to g^{*I}$ is taken  \textit{after} differentiation w.r.t. to the couplings.} The latter are introduced through \eqref{eq:BoundCouplingDef}.
The beta functions of boundary couplings are derived by renormalizing connected $n-$point functions of fundamental fields using the usual textbook minimal subtraction (MS) arguments. Note that if the bulk-boundary propagator $\Delta(x, y_\parallel)$ of fundamental fields is non-zero the singularities of boundary couplings induce new UV divergences in the \textit{bulk} correlators
 \begin{equation}
\vev{\phi(x_1) \dots \phi(x_n)} \ni \frac{1}{\epsilon} \int d^{d-1} z_{\parallel} \Delta(x_1, z_\parallel) ... \Delta(x_n, z_\parallel) \; .
 \end{equation}

\subsection{Cuttoff regularization} \label{sec:CutoffReg}

Let us consider again a bulk two-point function of two identical bulk scalars $O$, where one of the operators is put at a small distance $\frac{1}{\Lambda}$ from the boundary
\begin{equation} \label{eq:OOcutoff}
\vev{O(x)O_\Lambda(y_\parallel)} \equiv \vev{O(x)O(y_\parallel, \left|y_\perp\right|= 1/\Lambda)} \; .
\end{equation}
Here $\Lambda$ should be interpreted as a UV-cutoff corresponding to the thickness of the boundary and we divide the correlator by a bulk factor $\Lambda^{ \gamma_O}$ to take care of bulk divergences mentioned in \eqref{eq:BulkOren}. Using the BOE formula \eqref{eq:BOEcft} we have that
\begin{equation} 
\begin{aligned}
O_\Lambda(y_\parallel) &= \sum_{\hat{O}}\mu^O{}_{\hat{O}}\hat{O}(y_\parallel) \Lambda^{\D - \hD} \ ,
\end{aligned}
\end{equation}
from which we see that divergences arise from boundary operators with scaling dimension $\hat{\Delta}< \Delta$. 
In a usual bCFTs we only expect finitely many of such operators so to simplify the argument we will look at the case when there is only one. I.e.
\begin{equation} \label{eq:Oboe}
O_\Lambda(y_\parallel) = \mu^O{}_{\hat{O}} \hat{O}(y_\parallel) \Lambda^{\D - \hD} + \dots \ , 
\end{equation}
where $\dots$ stand for terms that vanish in the $\Lambda \to \infty$ limit. The corresponding divergent structures in the two-point correlator can be identified by substituting \eqref{eq:Oboe} in \eqref{eq:OOcutoff}.   Assuming the theory is conformal we can further apply \eqref{eq:BoundaryCorrelator} to obtain
\begin{equation}\label{eq:DivergentCorr}
\vev{O(x)O_\Lambda(y_\parallel)} =  \left(\frac{\Lambda}{\mu}\right)^{\D - \hD} \frac{ \mu^{\D - \hD} (\mu^O{}_{\hat{O}})^2 }{x_\perp^{\D - \hD}(s^2+x_\perp^2)^{ \hD}} + \dots \; ,
\end{equation}
where we have introduced an arbitrary IR scale $\mu$ to keep track of mass dimensions. 
 The above can therefore serve as a map from bulk-bulk correlator divergences to renormalized bulk-boundary correlators. In this case the renormalized correlator is obtained by dividing the l.h.s. of \eqref{eq:DivergentCorr} by $Z= \left(\frac{\Lambda}{\mu} \right)^{\D - \hD}$ and taking the $\Lambda \to \infty$ ($y_\perp \to 0$) limit.  Note that in order to derive \eqref{eq:DivergentCorr} we had to \textit{assume} the form \eqref{eq:BoundaryCorrelator} as the \textit{definition} of renormalized correlators. In particular different choices of the normalization of $\vev{\hat{O}\hat{O}}$ would lead to different \eqref{eq:BoundaryCorrelator} and therefore different renormalized $\hat{O}$. In the RG language this ambiguity corresponds to a choice of scheme, which is encoded in the finite part of the $Z-$factors.
 
\quad\quad In principle we could repeat the $Z-$ factor analysis of the previous section but in practice (especially when there are multiple operators) it easier to read-off the conformal data on $\hat{O}$ from the coefficient of the divergent term in \eqref{eq:DivergentCorr}. In general there will also be further divergent $s_\para^2-$dependent terms from descendants, but these do not contain any new data as they can be resumed following the arguments given in  Appendix \ref{app:BulkBdyCorr}.

 \quad\quad In the perturbation theory \eqref{eq:DivergentCorr} turns into either power divergence or logarithmic divergence depending on the scaling dimensions $O, \hat{O}$.  The individual boundary operators in the BOE can be identified by taking a boundary limit of suitable bulk operators in the free-field theory. When we include the (small) interactions both BOE coefficients and dimensions of the operators receive corrections. For example if $\Delta^{free}=\hat{\Delta}^{free}$ the divergence in \eqref{eq:DivergentCorr} becomes logarithmic 
 \begin{equation}\label{eq:DivergentCorrPert}
\vev{O(x)O_\Lambda(y_\parallel)} \approx  (\gamma_{O} - \gamma_{\hat{O}} )\log \left(\frac{\Lambda}{\mu}\right)\frac{  (\mu^O{}_{\hat{O}}^{(0)})^2 }{(s^2+x_\perp^2)^{ \Delta}} + finite \; ,
\end{equation}
where $\gamma$s represent the respective anomalous dimensions and $\mu^O{}_{\hat{O}}^{(0)}$ is the leading (free) BOE coefficient. Note that this logarithmic divergence would appear even if $\D< \hD$.
In general the leading order behaviour of the BOE in perturbation theory should look like
  \begin{equation} \label{eq:PertBOE}
 O \stackrel{x_\perp \to 0}{\approx} \sum_{\Delta^{free}>\hat{\Delta}_i^{free}} \mu^{O}{}_{i} \Lambda^{\Delta-\hat{\Delta}_i} \hat{O}_i + \mu^{O}{}_{\hat{O}}\left(1+ (\gamma_O - \gamma_{\hat{O}}) \log \Lambda \right) \hat{O} + \mathcal{O}\left(\frac{1}{\Lambda}\right)\; ,
 \end{equation}
where the summed term involves polynomial divergences that come from operators of lower engineering dimension as $O$ and the second term comes from the operators with $\gamma$s denoting the respective anomalous dimensions. The power divergences arise in perturbation theory when we close any two legs of $O_n$ which contributes a factor of $\frac{1}{x_\perp^{d-2}}$. On the Lagrangian level they correspond to additive renormalization of the sources 
\begin{equation} \label{eq:CutoffCount}
\int  d^{d-1} x_\parallel  \Lambda^{d-1 - \hat{\Delta}_i } \hat{O}_{i} \quad etc. \; ,
\end{equation}
which is similar to the way they appear in effective field theories from closing loops of massive UV fields.

\subsubsection{Example: anomalous dimensions from bootstrap} \label{sec:BulkModel}

As a warm-up example of the above method let us discuss the $\ph^4$-theory at its WF fixed point\footnote{Here $\mathbb{R}^d_+ = \mathbb{R}^{d - 1}\times\mathbb{R}_ {\geq 0} \ .$}
\begin{equation}
\begin{aligned}
S = \int_{\mathbb{R}^d_+}d^dx \left( \frac{(\p_\mu\phi^i)^2}{2} + \frac{g}{4!}(\phi^i)^4 \right) \ , \quad d = 4 - \e \ ,
\end{aligned}
\end{equation}
where that scalar $O(N)$ indices run from $1$ to $N$ and we used a notation $(\ph^i)^4 \equiv (\ph^2)^2$.

\quad \quad We equip the boundary with either Neumann or Dirichlet boundary condition
\begin{equation}
\begin{aligned}
\lim_{x_\perp \to 0}\phi^i &= 0 \ , \quad \text{(Dirichlet) ,} \\
\lim_{x_\perp \to 0} \p_\perp\phi^i &= 0 \ , \quad \text{(Neumann) .}
\end{aligned}
\end{equation}

This theory is has a WF fixed point where it is conformal at
\begin{equation}
\begin{aligned}
g &= \frac{48\pi^2\e}{N + 8} + \mco(\e^2) \ .
\end{aligned}
\end{equation}

The equations of motion are
\begin{equation} \label{eq:EoM}
\begin{aligned}
\p^2\ph^i &= \frac{g}{3!}\left(\ph^3\right)^i \ , \quad \left(\ph^3\right)^i \equiv \left(\ph^j\right)^2\ph^i \ .
\end{aligned}
\end{equation}

The $\ph-\ph$ correlator in this theory was found in \cite{1808.08155} through conformal bootstrap (up to an overall normalization constant)\footnote{Here and for the rest of this section, we will neglect the overall factor of $\de^{ij}$ in the correlators.}
\begin{equation} \label{eq:PhiPhi}
\begin{aligned}
\langle\ph(x)\ph(y)\rangle &= \frac{(\x + 1)^{\g - \hat{\g}}}{\left|x - y\right|^{2\D}} \pm \frac{\x^{\g - \hat{\g}}}{\left|\tilde{x} - y\right|^{2\D}} + \mco(\e^3) \ , \quad \x = \frac{(x - y)^2}{4x_\perp y_\perp} \ .
\end{aligned}
\end{equation}
Here $\pm$ denotes Neumann/Dirichlet boundary condition, $\x$ is the conformal cross-ratio, $\D$ is the full scaling dimension of $\ph$
\begin{equation}
\begin{aligned}
\D &= \frac{d - 2}{2} + \g \ , \quad \g &= \frac{N + 2}{4(N + 8)^2}\e^2 + \mco(\e^3) \ ,
\end{aligned}
\end{equation}
and $\g$ and $\hat{\g}$ are the anomalous dimensions of $\ph$ as well as the lowest dimensional primary that appears in the BOE of $\ph \ .$

\quad \quad For Neumann boundary conditions $\hat{\g}$ is the anomalous dimensions of $\hp$
\begin{equation}
\begin{aligned}
\hD_N &= \frac{d - 2}{2} + \hat{\g}_N \ , \quad \hat{\g}_N &= -\frac{N + 2}{2(N + 8)}\e - \frac{5(N + 2)(N - 4)}{4(N + 8)^3}\e^2 + \mco(\e^3) \ ,
\end{aligned}
\end{equation}

and for Dirichlet boundary condition, it is the anomalous dimension of  $\p_\perp\hp$
\begin{equation}
\begin{aligned}
\hD_D &= \frac{d}{2} + \hat{\g}_D \ , \quad \hat{\g}_D &= -\frac{N + 2}{2(N + 8)}\e - \frac{(N + 2)(17N + 76)}{4(N + 8)^3}\e^2 + \mco(\e^3) \ .
\end{aligned}
\end{equation}

The leading divergent parts of (\ref{eq:PhiPhi}) are
\begin{equation}
\begin{aligned}
\langle\ph(x)\ph_{\Lambda}(y_\parallel)\rangle &= \frac{2}{\left(s_\para^2 + x_\perp^2\right)^{\D}} \left( \frac{x_\perp}{4} \right)^{\g - \hat{\g}} \Lambda^{\g - \hat{\g}} \ , \\
\langle\ph(x)\p_\perp \ph_{\Lambda}(y_\parallel)\rangle &= 4 \left( 1 - \g + \hat{\g} \right)\frac{ \left( \g - \hat{\g} \right)s^2 + \left( \D + \g - \hat{\g}\right)x_\perp^2 }{x_\perp\left(s_\para^2 + x_\perp^2\right)^{\D +1}} \left( \frac{x_\perp}{4} \right)^{\g - \hat{\g}} \Lambda^{\g - \hat{\g}}  \ .
\end{aligned}
\end{equation}
Note that in the in the correlator on the second line we first evaluated the normal derivative and only then introduced the cutoff. These correlators have exactly the form \eqref{eq:DivergentCorr}.\footnote{We can absorb the finite factor $\left(\frac{1}{4}\right)^{\g - \hat{\g}}$ to the definition of BOE coefficients.} From this we find that the BOE of $\ph$ and $\p_\perp\ph$ is
\begin{equation}
\begin{aligned}
\ph_{\Lambda}(y_\parallel) &=  \Lambda^{ \g-\hat{\g}}\hp(y_\para) \ , \\
\p_\perp\ph_{\Lambda}(y_\parallel) &=  \Lambda^{ \g-\hat{\g}}\p_\perp\hp(y_\para) \ ,
\end{aligned}
\end{equation}
where we have abused the notation by defining $\p_\perp\hp(y_\para)$, which \textit{should not} be understood as an actual derivative of $\hp$. 

\quad \quad As a side note let us remark that the form of two-point function in \eqref{eq:PhiPhi} is partially dictated by the image symmetry $(x \leftrightarrow \tilde{x})$\footnote{One can show that the symmetry has to hold at all orders in perturbation theory, provided no boundary couplings are present.} which implies that the two-point function has to be of the form 
\begin{equation} \label{eq:PhiPhiGen}
\begin{aligned}
\langle\ph(x)\ph(y)\rangle &= \frac{f(\x + 1)}{\left|x - y\right|^{2\D}} \pm \frac{f(\x)}{\left|\tilde{x} - y\right|^{2\D}} \; ,
\end{aligned}
\end{equation}
where $f$ is an arbitrary function. The boundary limit corresponds to $\x \sim \Lambda \to \infty$ and we know from the free field limit of that at the boundary there exist renormalized scalars $\hp$, $\partial_\perp \hp$ of the same engineering dimensions as their bulk counterparts. The RG argument then implies that for large $x$ the function behaves as $f(x) \propto x^{\g-\hat{\g}}$, where  $\hat{\g}$ is the scaling dimension of either $\hp$ or $\partial_\perp \hp$. This argument only fixes the leading, divergent part of $f$, but in general there will also be finite, subleading contributions (the dots in \eqref{eq:DivergentCorr}) which correspond to boundary operators of higher dimension. More information about these terms can be obtained by taking normal derivatives of \eqref{eq:PhiPhiGen}, which turns the leading logarithmic divergences into power divergences and the subleading terms of the form $y_\perp^n \log y_\perp$ into logarithmic divergences. Since the the BOE coefficients of these operators start at $\mathcal{O}(\e)$ from \eqref{eq:DivergentCorr} we see their contribution to the correlator starts at $\mathcal{O}(\e^2)$ \cite{1808.08155}. Thus at order  $\mathcal{O}(\e^2)$ the only source of non-analyticity (logarithms) is the contribution of  $\hp$, $\partial_\perp \hp$, which explains why $f$ is uniquely determined from the leading divergence. At  $\mathcal{O}(\e^3)$ the anomalous dimensions of these higher dimensional operators start contributing and that changes the form of $f$. Before we move on let us determine, what operators they actually are.

\begin{itemize}
	\item{\textit{Neumann Boundary Condition}} \\
	The first subleading operator of engineering dimension $3$ is determined by taking two normal derivatives of \eqref{eq:PhiPhi}. By using equations of motion (\ref{eq:EoM}) in the boundary limit 
	\begin{equation} \label{eq:BulkEoM}
	\begin{aligned}
	\p_\perp^2\phi_{\Lambda}^i =  \p_\para^2\phi_{\Lambda}^i - \frac{g}{3!}(\phi_\Lambda^3)^i\ .
	\end{aligned}
	\end{equation}
	we identify $\phi^3$ as the origin of the operator we are looking for since the operator $\p_\para^2\phi$ corresponds to a descendant of $\hp$. The factor of $g$ is consistent with the observation of \cite{1808.08155} that the BOE coefficient ${\mu_{}^{\phi}}_n$ starting at $\mathcal{O}(\e)$.\footnote{Please note that from conformal bootstrap one actually finds the BOE coefficients squared.} More precisely the BOE of\footnote{In this BOE the cutoff is shifted by a factor of four, i.e. $\La \equiv (4y_\perp)^{-1}$.}
	\begin{equation} \label{eq:P3boe}
	(\phi_\Lambda^3)^i_\La(x_\para) = 16\e \left( \g - \hat{\g} + \al^2\e \right) \frac{\La^{2 + \al\e}}{\m^{\al\e}} \hp^i(x_\para) + \left( \frac{\La}{\m} \right)^{\g - \g_{\hp^3}} (\hp^3)^i(x_\para) \; , 
	\end{equation}
	
	where $\al = \frac{N + 2}{2(N + 8)}$ and the power divergence comes from closing two external legs of $\phi^3$, whereas the second term involves a boundary primary denoted as $(\hp^3)$ with anomalous dimension $\gamma_{\hp^3}$.\footnote{The actual effect of  $\gamma_{\hp^3}$ on $\vev{\phi \phi}$ will only appear at $\mathcal{O}(\e^3)$.}
The appearance of primary operator 	$(\hp^3)$ is purely an effect of adding a boundary as the equation of motion implies that $\phi^3$ is a descendant of $\phi$ in the absence of a boundary.\footnote{This also implies that the bulk anomalous dimension of $\phi^3$ is the same as $\phi$}
	The power divergence in \eqref{eq:P3boe} exactly corresponds to the one we get from differentiating the logarithmic divergence of $\phi$ twice so the new operator we are looking for is $(\hp^3)^i$. The anomalous dimension of this operator can be found by computing the two-point function of $\phi^3$ at $\mathcal{O}(\epsilon)$ and extracting the logarithmic divergence, whereas the the corrections to BOE coefficient follow from the $\vev{\phi \phi^3}$ correlator (we will describe how such computations are done for a cousin of this model in section \ref{sec:PhiSquared}). 
	
\quad \quad 	 To find the operators of dimension $5$ and higher we can take further normal derivatives of \eqref{eq:BulkEoM} and iterate the equation of motion. Doing this we find that the operators contributing to $\mathcal{O}(\e^2)$ are of the form  $\partial_{\parallel}^{2n} \hp^3$ for $n \geq 1$, where $\partial_{\parallel}^{2n}$ stands for all the terms with various redistributions of the derivatives. These operators are indistinguishable at $\mathcal{O}(\epsilon^2)$ as far as their contribution to the BOE goes. At $\mathcal{O}(\epsilon^3)$ they will all contribute to logarithmic divergences in the derivatives of $\vev{\phi \phi}$ through their anomalous dimensions (cf. \eqref{eq:DivergentCorrPert}). 
	 The operators of the form $\left(\hp^{2n + 1}\right)$ for $n \geq 2$ will appear only at  $\mathcal{O}(\e^3)$ and beyond as they will involve more powers of the coupling from the equation of motion.
	
	\item{\textit{Dirichlet Boundary Condition}} \\
	The case with Dirichlet boundary conditions is similar to that the Neumann case. The only difference is that the boundary condition now eliminates all the terms without normal derivatives (e.g. $\hp^3$ etc.) so we need to differentiate \eqref{eq:BulkEoM} at least three more times to obtain the first non vanishing operator, which is $(\p_\perp \hp)^3$. Thus we conclude that in the Dirichlet case the new operators start contributing from dimension $6$ and higher, which is exactly of what was observed in \cite{1808.08155}. 

\end{itemize}

\subsection{The boundary stress-energy tensor} \label{sec:SEtensorGeneral}

Let us now briefly turn to the question of renormalization of the SE tensor in the presence of boundary. In classic works \cite{Brown:1979pq, Brown:1980qq} it was show that the SE tensor does not undergo multiplicative renormalization in dimreg.\footnote{Nevertheless it still has to be renormalized additively to remove the power divergences (e.g. the cosmological constant) in schemes which break conformal invariance.} I.e.
\begin{equation} \label{eq:BulkTenRen}
T_{\mu \nu}= [T_{\mu \nu}]
\end{equation}

in the notation of Section \ref{sec:Divergences}. This is expected since $T_{\mu \nu}$ yields conserved Noether charges of spacetime symmetries and thus cannot depend on the RG scale. The boundary breaks translational invariance so one might expect some new divergence to arise when the boundary limit of \eqref{eq:BulkTenRen} is taken. This also means that the boundary conditions \eqref{eq:MomentumCond} and \eqref{eq:DisplacementDef} have to be understood as relations between bare operators in the language of Section \ref{sec:Divergences}.

\quad\quad To start with, let us analyse \eqref{eq:MomentumCond}, which imposes the conservation of the momentum parallel to the boundary. Here the operator $\hat{\tau}^{ab}$ can be thought of as the SE tensor for the boundary degrees of freedom. Its non-conservation means that there is a flow of energy/momentum from the bulk to boundary and vice versa. It has been recently argued by explicit computation in  mixed dimensional QED \cite{DiPietro:2019hqe} that unlike its bulk counterpart \eqref{eq:BulkTenRen} it can acquire anomalous dimension. The appearance of such anomalous dimension implies that the BOE of $T^{a \perp}$ takes the form
\begin{equation} \label{eq:TnaBOE1}
T^{a \perp} \stackrel{x_\perp \to 0}{\sim} {\mu_T}^{\hat{\tau}}x_\perp^{\Delta_{\hat{\tau}^{ab}}-d+1} \partial_b \hat{\tau}^{ab}+ \dots \; .
\end{equation}
Assuming that $\hat{\tau}^{ab}$ has the properties of a spin $2$ operator at the boundary we have that $\Delta_{\hat{\tau}^{ab}}>d-1$ in a unitary theory. In particular this implies vanishing of the l.h.s. of \eqref{eq:TnaBOE1} in the boundary limit in accordance with \eqref{eq:CardyCon}.\footnote{The authors would like to thank Davide Gaiotto for explaining this point.}

\quad\quad In this paper we will consider a setup described in \eqref{eq:BoundCouplingDef} where the bulk theory is deformed by a (nearly-)marginal scalar operator(s) at the boundary. For simplicity we will consider a single coupling case  with a classically marginal operator $O_{d-1}$ and the corresponding coupling $\hat{g}$. Based on dimensional grounds and symmetry we expect that
\begin{equation} \label{eq:TnaBOE2}
\lim_{x_\perp \to 0} T^{a \perp} \propto \partial^a O_{d-1} \; .
\end{equation}
The argument of \eqref{eq:TnaBOE1} is now complicated by the fact the theory is not conformal and the unitarity bound for scalar is  below $(d-1)$. Since the violation of conformal invariance is controlled by the $\beta-$function of $g$ based on similarity with the trace anomaly in a usual QFT \cite{Collins:1976yq} we postulate that 
\begin{equation} \label{eq:TnaBOE3}
\lim_{x_\perp \to 0} T^{a \perp} \sim \beta \partial^a \hat{O}_{d-1}\; ,
\end{equation}
where the r.h.s. involves only manifestly finite, renormalized quantities and satisfies \eqref{eq:CardyCon} at the fixed points of $\beta$. By comparing \eqref{eq:MomentumCond} and \eqref{eq:TnaBOE3} we get that
\begin{equation}
\partial_b \hat{\tau}^{ab} \sim \beta \partial^a \hat{O}_{d-1} \ .
\end{equation}
Just as in the case of bulk trace anomaly, $\hat{\tau}^{ab}$ does not acquire any anomalous dimension provided the product $\beta \hat{O}_{d-1} $ is RG-invariant.
Let us now use some general RG arguments to study the behaviour \eqref{eq:TnaBOE3} in the IR regime where
\begin{equation} \label{eq:IRlimit}
\frac{1}{\mu} \gg x_\perp  \; .
\end{equation}
One expects that the BOE should be applicable in this regime if all other specific distance scales (here denoted by $\frac{1}{\mu}$) are much larger $x_\perp$.
If a theory has an interacting IR fixed point we have that
\begin{equation}
\beta(\mu) \sim (\mu x_\perp)^{\delta>0} \quad \text{for} \quad \mu x_\perp \ll 1 \; ,
\end{equation}
which yields exactly \eqref{eq:TnaBOE1} when substituted in \eqref{eq:TnaBOE3}.\footnote{In case of non-interacting IR fixed point we would get a logarithmic suppression since $\beta(\mu) \sim \frac{1}{\log(\mu x_\perp)}$.} The study of UV regime is more complicated due to the competition between small distance scales $\frac{1}{\mu}$ and $x_\perp$. 

\quad\quad At last let us comment on possible power divergences that might appear in the BOE \eqref{eq:TnaBOE1}. Given the fact the the boundary cutoff introduced in Section \ref{sec:CutoffReg} breaks explicitly both the conformal symmetry and normal diffeomorphisms one might expect polynomial divergences contributing to both $\hat{\tau}^{ab}$ and $\hat{D}$. As discussed around \eqref{eq:CutoffCount} they are to be canceled by fine-tuning the mass parameters.
A similar issue arises if one tries to renormalize the bulk SE tensor with a cutoff regulator in the presence of scalar fields (see for example \cite{Meissner:2007xv}). One can avoid this problem by using dimreg as outlined in Section \ref{sec:Dimreg}. Based on the usual argument we do not expect such power divergences to appear in supersymmetric theories.

\section{Boundary interaction} \label{sec:BoundaryExample}
Let us now introduce the model studied in this paper. A similar model involving $\mathbb{Z}_2-$ breaking cubic boundary  term in $d=4- \epsilon$ was considered in \cite{PhysRevB.44.6642}. A supersymmetric version of related models in $3d$ was recently analysed in \cite{Brunner:2019qyf}. Here we would like to consider an $O(N)-$invariant and non-supersymmetric theory with a free scalar CFT in the bulk perturbed by a marginal self-coupling at the boundary. The advantage of this model is that the bulk anomalous dimensions vanish so we do not have to worry about bulk renormalization effects. The relevant Lagrangian reads
\begin{equation} \label{eq:BdaryLagrangian}
\begin{aligned}
S = \int_{\mathbb{R}^d_+} d^dx \left( \frac{(\p_\mu\phi^i)^2}{2} + \de(x_\perp)\frac{\hg_0}{4!}(\ph^i)^4 \right) \ ,
\end{aligned}
\end{equation}
where that scalar $O(N)$ indices run from $1$ to $N$ and we used a notation $(\ph^i)^4 \equiv (\ph^2)^2$. Please remember from \eqref{eq:BoundCouplingDef} that the boundary limit of a bulk field in the corresponding coupling should be seen as a bare quantities. From dimensional analysis we find
\begin{equation} \label{eq:DimCoupling}
\begin{aligned}
\D = \hD = \frac{d - 2}{2} \quad\Rightarrow\quad \dim(\hat{g}_0) = 3 - d = \e \ , \quad \text{if $d = 3 - \e \ .$}
\end{aligned}
\end{equation}

This tells us that we should consider spacetime dimensions close to three if we wish to have a CFT on the boundary as well. We find the equations of motion by varying the action with respect to both $\ph(x_\perp \ne 0)$ and $\ph(x_\perp = 0)$, where we need to take into account boundary terms whenever we perform a partial integration. This gives us the Klein-Gordon equation of motion as well as a modified Neumann boundary condition
\begin{equation} \label{eq:ModifiedBC}
\begin{aligned}
\p^2\ph^i = 0 \ , \quad \lim\limits_{y_\perp\rightarrow 0}\p_\perp\ph^i =\lim\limits_{y_\perp\rightarrow 0} \frac{\hat{g}_0}{3!}(\ph^3)^i \ .
\end{aligned}
\end{equation}

This boundary condition can also be written as
\begin{equation} \label{eq:ModifiedBC2}
\begin{aligned}
\lim\limits_{y_\perp\rightarrow 0}\p_\perp\ph^2 = \lim\limits_{y_\perp\rightarrow 0}\frac{\hat{g}_0}{3}\ph^4 \ , \quad \ph^2 \equiv (\ph^i)^2 \ , \quad \ph^4 \equiv (\ph^2)^2 \ .
\end{aligned}
\end{equation}

In the free theory, the bulk-bulk correlators can be found using method of images
\begin{equation} \label{eq:phi-phiCorr}
\begin{aligned}
\langle\ph^i(x)\ph^j(y)\rangle &= A_d \delta_{ij}\left( \frac{1}{\left|x - y\right|^{2\D}} + \frac{1}{\left|\tilde{x} - y\right|^{2\D}} \right) \ , \quad A_d \equiv \frac{1}{(d - 2)S_{d - 1}} \ .
\end{aligned}
\end{equation}

Here $\tilde{x}$ is the image point of $x \ ,$ and $S_{d - 1}= \frac{2 \pi^{d/2}}{\Gamma_{d/2}}$ is the area of a $(d - 1)$-dimensional sphere. Let us point out an interesting fact that this correlator satisfies something called image symmetry, which means that if we reflect one of the coordinates in the boundary $x\leftrightarrow\tilde{x} \quad \Leftrightarrow x_\perp\rightarrow - x_\perp$ we get back the same correlator. From this bulk-bulk correlator can find the bulk-boundary and the boundary-boundary correlator by taking the boundary limit of the bulk operators
\begin{equation} \label{eq:BulkBdaryPP}
\begin{aligned}
\langle\ph^i(x)\hp^j(y)\rangle &= \lim\limits_{y_\perp\rightarrow 0}\langle\ph^i(x)\ph^j(y)\rangle = \frac{2A_d \delta_{ij}}{\left(s_\para^2 + x_\perp^2\right)^{2\D}} \ , \\
\langle\hp^i(x)\hp^j(y)\rangle &= \lim\limits_{x_\perp, y_\perp\rightarrow 0}\langle\ph^i(x)\ph^j(y)\rangle = \frac{2A_d \delta_{ij}}{\left|s_\para\right|^{2\D}} \ .
\end{aligned}
\end{equation}

The renormalized one-point function of $\ph^2$ is the finite part of (\ref{eq:phi-phiCorr}) in the coincident limit
\begin{equation}
\begin{aligned}
\langle\ph^2(x)\rangle &= \text{finite}\lim\limits_{x\rightarrow y}\langle\ph^i(x)\ph^i(y)\rangle = N \frac{A_d}{\left(2x_\perp\right)^{2\D}} \ .
\end{aligned}
\end{equation}

In three dimensions
\begin{equation} \label{eq:Const3Dim}
\begin{aligned}
A_3 = \frac{1}{4\pi} \ .
\end{aligned}
\end{equation}
In the following two section we will proceed to discuss the renormalization of the Lagrangian defined by \eqref{eq:BdaryLagrangian} in dimreg. Since the model has no bulk interaction all the renormalization constants are related strictly to boundary fields. In fact we will only have two independent constants: $Z_{\hg}$ for the coupling and the field renormalization defined through
\begin{equation}
\lim_{x_\perp \to 0} \phi^i(x) \equiv \sqrt{Z_{\hat{\phi}}} \hat{\phi}^i(x_\parallel) \; .
\end{equation}

\subsection{$\ph - \ph$ correlator} \label{sec:Phi}
\begin{figure} 
	\centering
	\includegraphics[width=0.25\textwidth]{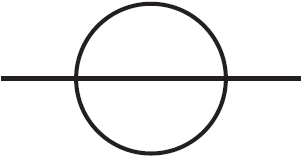}
	\caption{The $\ph-\ph$ correlator at order $\hg^2 \ .$}
	\label{Fig:phi-phi}
\end{figure}

We now proceed to compute the quantum corrections to the propagator \eqref{eq:phi-phiCorr} in the presence of non-zero boundary coupling $\hat{g}$. Our strategy will be to use dimreg and compute the correlator in momentum space. In dimreg the leading (power-divergent) diagram vanishes and we are left with the diagram on Figure \ref{Fig:phi-phi} at $\mathcal{O}(\hg^2)$.  
 In principle one could try to evaluate the correlator in position space with boundary cutoff in place and then subtract the power divergences by adding a boundary mass term
\begin{equation} \label{eq:PowerDivPP}
\int d^{d-1} x_\parallel \Lambda \phi^2 \; .
\end{equation}

This divergent term resembles the radiative corrections scalar mass in effective field theories and therefore could have an interesting physical interpretation in general. At next order the diagram on Figure \ref{Fig:phi-phi} will have both power divergences of the type \eqref{eq:PowerDivPP} and short-distance UV divergences in $\epsilon$.\footnote{The authors would like to thank Pierre Vanhove for pointing this out and explaining how the techniques used in \cite{Bloch:2013tra} can be used to tackle this problem.} We will leave the investigation of power divergences for the future.

\quad\quad As already mentioned, we can avoid power divergences by Fourier transforming and computing the diagram in momentum space (cf. Footnote \ref{Footnote:PowerDiv}). The bulk-boundary propagator in momentum space is obtained by Fourier transforming \eqref{eq:BulkBdaryPP}
 \begin{equation} \label{eq:MomSpaceProp}
G^{ij}(\underline{p}, x_\perp) \equiv \int d^{d - 1} s_\parallel e^{i \underline{p} \cdot  \underline{s}_\parallel} \langle\ph^i(x)\hp^j(y_\para)\rangle   = \oo_d \frac{ e^{- |\underline{p}|x_\perp}}{\left|\underline{p}\right|}\de^{ij} \ .
\end{equation}

Here $\underline{p}$ is momentum parallel to the boundary and the overall factor is
\begin{equation}
\begin{aligned}
\oo_d \equiv 2^{d - 2\D}\pi^{(d - 1)/2}\frac{\G_{(d - 1)/2 - \D}}{\G_\D}A_d = \frac{2^{d - 2\D - 2}}{(d - 2)\sqrt{\pi}}\frac{\G_{d/2}\G_{(d - 1)/2 - \D}}{\G_\D} \ ,
\end{aligned}
\end{equation}

where we have denoted $\G$-functions as $\G(x) \equiv \G_x $. In three dimensions this factor simplifies significantly
\begin{equation}
\begin{aligned}
\D \to \frac{1}{2}  \quad\Rightarrow\quad \oo_3 = 1 \ .
\end{aligned}
\end{equation}

The all-order corrections to the free propagator in momentum space take a form
\begin{equation} \label{eq:2ptMom}
\D^{ij}(\underline{p}, x_\perp, y_\perp) = G^{ik}(\underline{p}, x_\perp) \Pi^{kl}(\underline{p}^2)G^{l j}(-\underline{p}, y_\perp) \; ,
\end{equation}

where we defined the self-energy $ \Pi^{kl}(\underline{p}^2)$, which only depends on $\underline{p}^2$ since the theory has no bulk couplings. By power counting the mass dimension of $\Pi^{kl}(\underline{p}^2)$ is one so it has to be finite in a massless theory. The reason for this is that a divergence would imply a counterterm of the form
\begin{equation}
\frac{1}{\epsilon}|\underline{p}| \stackrel{pos. \ space}{\to} \frac{1}{\epsilon}\int d^{d - 1}x_\parallel \phi^i\sqrt{\partial_\parallel^2} \phi^i \ ,
\end{equation}

which is clearly non-local. This means that the self-energy can only have sub-divergences corresponding to the coupling renormalization and we have that
\begin{equation} \label{eq:ZphiCond}
Z_{\hat{\phi}}= 1
\end{equation}

to all orders in perturbation theory. A similar observation has been made for the boundary $\phi^3$ theory in \cite{PhysRevB.44.6642}.  Furthermore if we tune to a fixed point $\hg \to \hg^*$ (details and existence of this fixed point will be explained in Section \ref{sec:BetaDeriv}) also the subdivergences cancel out and we are left with
\begin{equation}
\Pi^{kl}(\underline{p}^2)= C |\underline{p}| \delta^{kl}\; , 
\end{equation}
for some finite constant $C$. We can plug this expression back to \eqref{eq:2ptMom} to find the correction to the propagator which now takes a form
\begin{equation}
\D^{ij}(\underline{p}, x_\perp, y_\perp) = (\oo_d C) \oo_d \frac{ e^{- |\underline{p}|(x_\perp+y_\perp)}}{\left|\underline{p}\right|}\de^{ij} \; .
\end{equation}

Note that the above expression is exactly of the form \eqref{eq:MomSpaceProp} and thus we can use \eqref{eq:BulkBdaryPP} to find the form of all-order correction to \eqref{eq:phi-phiCorr}
\begin{equation}
\langle\ph^i(x)\ph^j(y)\rangle_{\text{correction}} = (2\oo_d C) \frac{A_d\de^{ij}}{\left|\tilde{x} - y\right|^{2\D}} \; .
\end{equation}

In another words the boundary interactions in a bCFT only change the reflected part of the correlator in \eqref{eq:phi-phiCorr}. Note that if we did not tune the coupling to a fixed point the coefficient $C$ would depend on momentum through the running coupling $\hg(\underline{p}^2)$ and the corresponding Fourier transform would involve more complicated dependence on the the normal coordinates.

\quad\quad We can compute the $\hg^2$ correction to $C$ by evaluating the amputated part of the diagram on Figure \ref{Fig:phi-phi} which reads
\begin{equation}
\Pi^{ij}(\underline{p}^2)= \delta^{ij} \frac{\hg^2}{18} (N+2)\oo_d^3 \int d^{d-1} \underline{k_1} \int d^{d-1}\underline{k_2}\frac{1}{|\underline{k}_1|}\frac{1}{|\underline{k}_2|}\frac{1}{|\underline{p}-(\underline{k}_1+\underline{k}_2)|} \; .
\end{equation}

This integral can be evaluated by standard methods (iterating the one-loop massless bubble) and yields a finite result as expected from \eqref{eq:ZphiCond}. The result of the computation reads
\begin{equation}
\Pi^{ij}(\underline{p}^2)= - \delta^{ij} (N+2)\frac{\hg^2}{72 \pi^2}|\underline{p}|+ \mathcal{O}(\hg^3) \; ,
\end{equation}

from which we can read off the correction to the reflection coefficient
\begin{equation} \label{eq:2loopCorrelator}
\langle\ph^i(x)\ph^j(y)\rangle = A_d\de^{ij}\left( \frac{1}{\left|x - y\right|^{2\D}} + \frac{1- (N+2)\frac{\hg^2}{36 \pi^2}}{\left|\tilde{x} - y\right|^{2\D}} \right) + \mathcal{O}(\hg^3) \;   .
\end{equation}

This bulk-bulk correlator has no divergences in the boundary limit, which means the boundary field $\hp^i$ receives no anomalous dimension and by the above argument
\begin{equation} \label{eq:phiNonRen}
\g_{\hp} = 0  \ 
\end{equation} 
to all orders in perturbation theory.\footnote{Note that in the conformal case this can be also seen as a consequence of equation of motion. We thank Marco Meineri for pointing this out. } Furthermore let us remark that by inspecting \eqref{eq:2loopCorrelator} we see that the image symmetry gets broken at $\mathcal{O}(\hg^2)$ and the one point function of $\phi^2$ receives a correction
\begin{equation}
\vev{\phi^2(x)} \equiv \lim_{x \to y} \langle\ph(x)\ph(y)\rangle = \frac{A_d}{4 x_\perp^{d-2}}\left(1- (N+2)\frac{\hg^2}{36 \pi^2}\right) + \mathcal{O}(\hg^3) \; .
\end{equation}

\subsection{Renormalization group flow of the theory and its fixed points} \label{sec:BetaDeriv}

Let us move on to find the $\be$-function for this coupling
\begin{equation}
\begin{aligned}
\be = \m\frac{d\hat{g}}{d\m} = \frac{d\hat{g}}{d\log(\m)} \quad\Rightarrow\quad \frac{\be}{\hat{g}} = \frac{d\hat{g}}{d\log(\m)}\frac{d\log(\hat{g})}{d\hat{g}} = \frac{d\log(\hat{g})}{d\log(\m)} \ .
\end{aligned}
\end{equation}

Here $\m$ is the renormalization scale and $\hat{g}$ is the renormalized coupling ($\hat{g}_0$ is the bare coupling). In dimreg we find how $\hat{g}$ depends on $\m$ by assuming that it does not matter whether we renormalize the fields or the coupling constant
\begin{equation} \label{eq:g0Def}
\hat{g}_0 = \left[Z_{\hat{\ph}}(\m)\right]^{-2} Z_{\hat{g}}(\m)\m^{\dim \hat{g}_0}\hat{g}(\m) \; ,
\end{equation}
where $\dim \hat{g}_0= \epsilon$ from \eqref{eq:DimCoupling}.
By definition, a bare coupling does not depend on the renormalization scale. The factor $Z_{\hat{\ph}}$ was found in the previous section \eqref{eq:ZphiCond}, while $Z_{\hat{g}}$ is found from the divergences of the amputated four-point function $\hp$. So to find the $\be$-function we only need to study scalar field diagrams on the boundary since the amputated graphs only involve boundary propagators.\footnote{This means the $\be$-function we find in this section might just as well be for a theory on an interface, where there are two (possibly different) scalar theories on each side of it.} If we differentiate with respect to $\log(\m)$
\begin{equation} \label{eq:Beta}
\begin{aligned}
\frac{\be}{\hat{g}} &= 2\frac{d\log(Z_\hp)}{d\log(\m)} - \frac{d\log(Z_{\hat{g}})}{d\log(\m)} - \dim\hat{g}_0 \ .
\end{aligned}
\end{equation}

The $Z_{\hg}$-factor is on the form (in the MS scheme)
\begin{equation} \label{eq:Renorm}
\begin{aligned}
Z_{\hg} = 1 + a_{\hg}\frac{\hg}{\e} + \mco(\hg^2) \ .
\end{aligned}
\end{equation}

Here $a_{\hg}$ is a finite constant. We find
\begin{equation}
\begin{aligned}
\log(Z_{\hg}) = \frac{a_{\hg}\hg}{\e} + \mco(\hg^2) \quad\Rightarrow\quad \frac{d\log(Z_{\hg})}{d\log(\m)} = \frac{a_{\hg}}{\e}\be + \mco(g^2) \ ,
\end{aligned}
\end{equation}

Using this, together with \eqref{eq:DimCoupling} and \eqref{eq:ZphiCond}, we find the $\be$-function from (\ref{eq:Beta})
\begin{equation} \label{eq:RealBeta}
\begin{aligned}
\be &= - \e\hg \left( 1 + \frac{a_{\hg}}{\e}\hg \right) + \mco(\hat{g}^3) \ .
\end{aligned}
\end{equation}

The easiest way to find the $\be$-function is in momentum space. The renormalized four-point correlator is\footnote{This is done in the standard QFT way, and we have put the details on this calculation in Appendix \ref{App:4PtCorr}.}
\begin{equation} \label{eq:RenormCorr}
\begin{aligned}
G_4^{ijkl}(\{p_i\}) &= \oo_d^2 \left( D^{ijkl}  - \frac{\oo_d^2\hat{g}}{3}D^{ijkl} - \frac{\oo_d^4\hat{g}^2}{36\pi} \left(F^{ijkl}_{12} + F^{ikjl}_{13} + F^{iljk}_{14} \right) \right) \ ,
\end{aligned}
\end{equation}
\begin{equation*}
\begin{aligned}
F^{ijkl}_{ab} &= E^{ijkl}\log\left|\frac{p_a + p_b}{\m}\right| \ ,
\end{aligned}
\end{equation*}

with the $O(N)$ tensor structure
\begin{equation}
\begin{aligned}
D^{ijkl} &= \de^{ij}\de^{kl} + \de^{ik}\de^{jl} + \de^{il}\de^{jk} \ , \\
E^{ijkl} &= (N + 2)\de^{ij}\de^{kl} + 2D^{ijkl} \ .
\end{aligned}
\end{equation}

We have renormalized the coupling constant with (see (\ref{eq:Renorm}))
\begin{equation} \label{eq:Z1Coupling}
\begin{aligned}
a_{\hg} &= \frac{(N + 8)\oo_d^2}{12\pi} = \frac{N + 8}{12\pi} + \mco(\e) \ .
\end{aligned}
\end{equation}

\begin{figure} 
	\centering
	\includegraphics[width=0.5\textwidth]{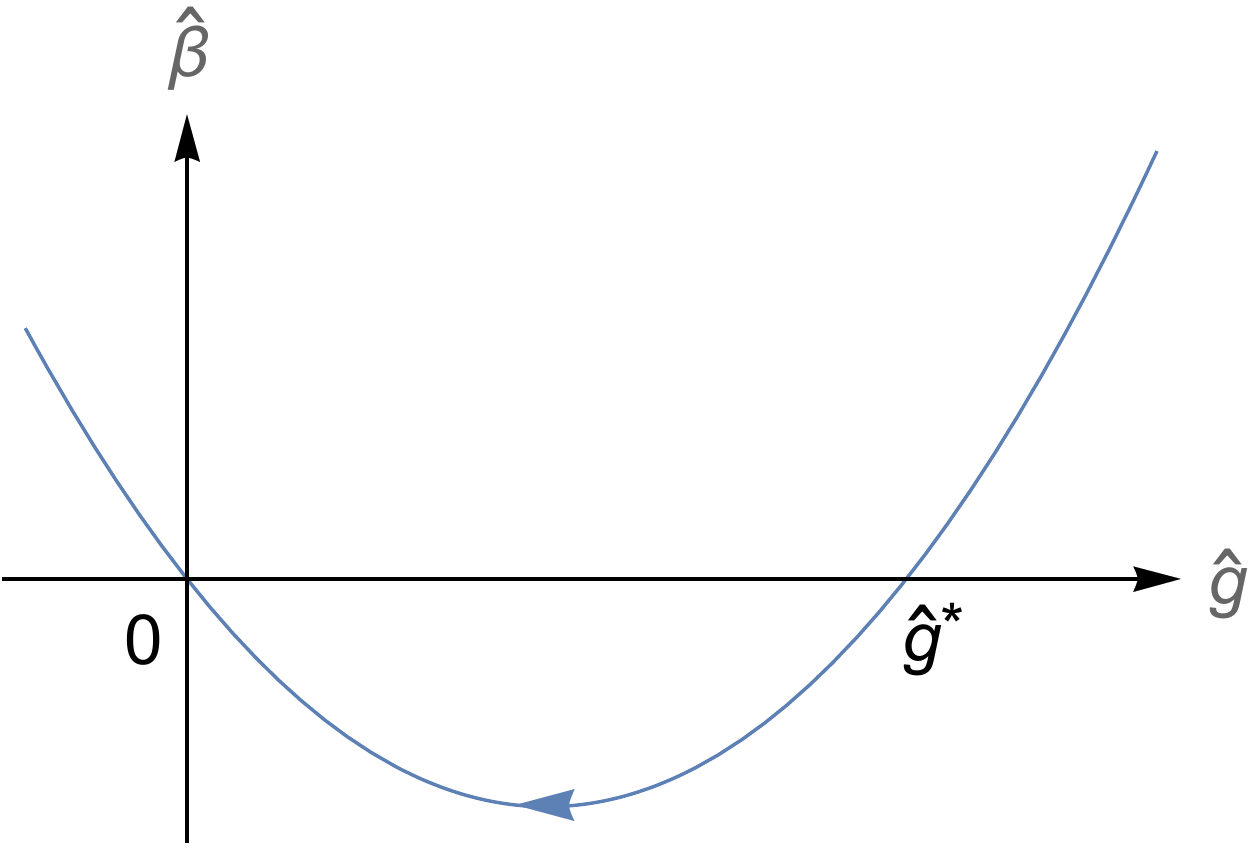}
	\caption{The RG flow of the theory for positive $\epsilon$, where $\hat{g}^* = \frac{12\pi\e}{(N + 8)\oo_d^2}$ and the arrow points from the IR to UV.} 
	\label{Fig:RGflow}
\end{figure}

If we insert this into (\ref{eq:RealBeta}) we find the $\be$-function
\begin{equation} \label{eq:FinalBeta}
\begin{aligned}
\be &= -\e\hat{g} + \frac{(N + 8)\oo_d^2}{12\pi}\hat{g}^2 + \mco(\hat{g}^3)  \ .
\end{aligned}
\end{equation}

From the above we conclude that the theory is asymptotically free for positive $\epsilon$. The RG flow described by this $\be$-function (cf. Figure \ref{Fig:RGflow}) has two fixed points where
\begin{equation} \label{eq:Be0}
\be = 0 \; .
\end{equation}

It has a Gaussian (UV) fixed point, but more interestingly also a WF (IR) fixed point at
\begin{equation} \label{eq:FixedPt}
\begin{aligned}
\hat{g}^* = \frac{12\pi\e}{(N + 8)\oo_d^2} + \mco(\e^2) = \frac{12\pi\e}{N + 8} + \mco(\e^2) \ .
\end{aligned}
\end{equation}
Let us now briefly discuss the nature of the fixed point \eqref{eq:FixedPt} as we vary the sign of $\epsilon$. We have essentially three possibilities:
\begin{itemize}
\item{$\epsilon>0$} {implies a UV-free theory with interacting IR fixed point. If one analytically continues to $\epsilon = 1$ this should correspond to a free bosonic theory in $2d$ with $\phi^4$ potential at the boundary.}
\item{$\epsilon=0$} {implies an IR-free theory by positivity of the $\mathcal{O}(\hg^2)$ term in \eqref{eq:FinalBeta}. Here the coupling runs into a Landau pole in the UV so one might need more couplings  or fields to find a UV complete description of the theory.  }
\item{$\epsilon<0$} {implies an IR-free theory with interacting fixed point in the UV. Going to $\e = -1$ it should correspond to a $4d$ scalar field theory with unstable potential at the boundary (sign of the quartic interaction is negative).  }
\end{itemize}

\subsection{$\ph^2 - \ph^2$ Correlator} \label{sec:PhiSquared}

The $\ph^2-\ph^2$ correlator is a good example to show that physical quantities, like anomalous dimensions, are independent of the renormalization scheme. We will study the divergences that appear in the boundary limit, and regularize them with dimreg as well as with a small distance cutoff. Please note that in the following computations we suppress the $O(N)$ notation. 
The connected part of the free theory correlator is\footnote{In this work we will not consider disconnected diagrams, which are actually more related to the one-point functions.}
\begin{equation} \label{eq:FreeThyCorr}
\begin{aligned}
\langle\ph^2(x)\ph^2(y)\rangle &= 2N\langle\ph(x)\ph(y)\rangle^2 = 2NA_d^2\left( \frac{1}{\left|x - y\right|^{2\D}} + \frac{1}{\left|\tilde{x} - y\right|^{2\D}} \right)^2 \ .
\end{aligned}
\end{equation}

The bulk-boundary and boundary-boundary correlator is found from the boundary limit of this correlator
\begin{equation} \label{eq:P2free}
\begin{aligned}
\langle\ph^2(x)\hp^2(y_\para)\rangle &= \frac{8NA_d^2}{\left(s_\para^2 + x_\perp^2\right)^{2\D}} \ , \\
\langle\hp^2(x_\para)\hp^2(y_\para)\rangle &= \frac{8NA_d^2}{|s_\para|^{2\D}} \ .
\end{aligned}
\end{equation}

\begin{figure} 
	\centering
	\includegraphics[width=0.5\textwidth]{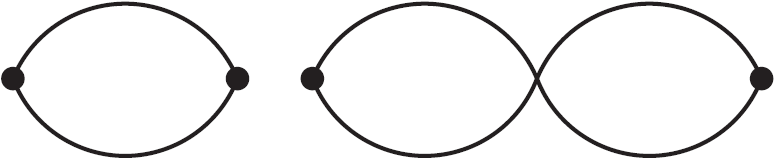}
	\caption{The $\ph^2 - \ph^2$ correlator up to order $\hg \ .$}
	\label{fig:Phi2diags}
\end{figure}

The $\ph^2-\ph^2$ correlator at order $\hg$ corresponds the diagrams on figure \ref{fig:Phi2diags}. The two-loop diagram contributes the following factor
\begin{equation} \label{eq:P2P21}
\begin{aligned}
\langle\ph^2(x)\ph^2(y)\rangle_1 &= -\frac{\hat{g}}{4!}8N(N + 2)\int d^{d - 1}z_\para\langle\ph(x)\hp(z_\para)\rangle^2\langle\hp(z_\para)\ph(y)\rangle^2 \\
&= -\frac{16N(N + 2)A_d^4\hat{g}}{3}I_{2\D, 2\D}^{d - 1} \ .
\end{aligned}
\end{equation}

Here $I_{\al\be}^n$ is the integral studied in Appendix \ref{app:ph^2Loop}.

\subsubsection{Dimensional regularization in the boundary limit} \label{sec:dimregPhi2}

We will now proceed as outlined in section \ref{sec:Dimreg} and \textit{first} take the $y_\perp \to 0$ limit of \eqref{eq:P2P21} and \textit{then} expand in epsilon. This yields the following boundary limits
\begin{equation} \label{eq:Corrs}
\begin{aligned}
\lim\limits_{y_\perp\rightarrow 0}\langle\ph^2(x)\ph^2(y)\rangle_1 &= -\frac{16\pi N(N + 2)A_d^4\overline{g}}{3\left(s_\para^2 + x_\perp^2\right)} \left( \frac{2}{\e} - \log(x_\perp^2) + 4\log(s_\para^2 + x_\perp^2) + \log(\m^2) \right) \ ,
\end{aligned}
\end{equation}
\begin{equation*}
\begin{aligned}
\lim\limits_{x_\perp, y_\perp\rightarrow 0}\langle\ph^2(x)\ph^2(y)\rangle_1 &= -\frac{32\pi N(N + 2)A_d^4\overline{g}}{3s_\para^2} \left( \frac{2}{\e} + 3\log(s_\para^2) + \log(\m^2) \right) \ . 
\end{aligned}
\end{equation*}

We wish to point out that these correlators satisfy image symmetry w.r.t. $x_\perp$.\footnote{Remember from Section \ref{sec:Phi} that image symmetry breaks down at order $\hg^2$.} Here we absorbed $\g_E$- and $\log(\pi)$-terms in a dimensionless $\overline{MS}$ coupling
\begin{equation} \label{eq:MSBarCoupling}
\begin{aligned}
\hat{g} &= \left(\pi e^{\g_E}\right)^{\e/2}\m^\e\overline{g} + \mco(\e^2) \ .
\end{aligned}
\end{equation}

In order to remove the divergence in the correlators which contain boundary fields, we have to renormalize the boundary limit of the bulk operator
\begin{equation} \label{eq:Zp2}
\begin{aligned}
\lim\limits_{x_\perp \rightarrow 0}Z_{\hat{\phi}^2}\ph^2 = \hp^2 \ , \quad Z_{\hat{\phi}^2} = 1 + a_{\hp^2}\frac{\overline{g}}{\e} + \mco(\tilde{g}^2) \ .
\end{aligned}
\end{equation}

This yields
\begin{equation*} 
\begin{aligned}
\langle\ph^2(x)\hp^2(y)\rangle &= \lim\limits_{y_\perp \rightarrow 0}Z_{\hat{\phi}^2}\langle\ph^2(x)\ph^2(y)\rangle \\
&= \frac{8NA_d^2}{s^2 + x_\perp^2} \left[ 1 + \frac{\overline{g}}{3} \left( \frac{3a_{\hp^2} - 4\pi(N + 2)A_d^2}{\e} + 2\pi(N + 2)A_d^2\log(x_\perp^2) + \rig\rig \\
\eq\lef\lef + \left( 3a_{\hp^2} - 8\pi(N + 2)A_d^2 \right)\log(s^2 + x_\perp^2) - 2\pi(N + 2)A_d^2\log(\m^2) \frac{}{}\right)\right] + \mco(\e^2, \overline{g}^2) \ .
\end{aligned}
\end{equation*}

We can now set $a_{\hp^2}$ such that the divergence disappears
\begin{equation} \label{eq:aPhi2Coeff}
\begin{aligned}
a_{\hp^2} = \frac{4\pi(N + 2)A_d^2}{3} \ .
\end{aligned}
\end{equation}

We can also determine the renormalized correlator $\vev{\hp(x_\para)\hp(y_\para)}$ from the second line of \eqref{eq:Corrs} whose divergence is absorbed by $(Z_{\hp^2})^2$.
In the MS scheme the renormalized correlators are
\begin{equation} \label{eq:BndyIntCorr}
\begin{aligned}
\langle\ph^2(x)\hp^2(y_\para)\rangle_{1} &= \frac{16\pi N(N + 2)A_3^4\overline{g}}{3}\frac{\log(x_\perp^2) - 2\log(s_\para^2 + x_\perp^2) - \log(\m^2)}{s_\para^2 + x_\perp^2} + \mco(\overline{g}^2) \ , \\
\langle\hp^2(x_\para)\hp^2(y_\para)\rangle_{1} &= -\frac{32\pi N(N + 2)A_3^4\overline{g}}{3}\frac{\log(s_\para^2) + \log(\m^2)}{s_\para^2} + \mco(\overline{g}^2) \ .
\end{aligned}
\end{equation}

By adding the above finite, one-loop results to \eqref{eq:P2free} and resumming we get
\begin{equation} \label{eq:BndyIntCorrRes}
\begin{aligned}
\langle\ph^2(x)\hp^2(y_\para)\rangle &= 8NA_d^2\frac{\mu^{-a_{\hp^2}\overline{g}} }{\left|x_{\perp}\right|^{-a_{\hp^2}\overline{g} }\left(s_\para^2 + x_\perp^2\right)^{2\D + a_{\hp^2}\overline{g}}} + \mco(\overline{g}^2) \ , \\
\langle\hp^2(x_\para)\hp^2(y_\para)\rangle_{1} &= 8NA_d^2\frac{\mu^{-2a_{\hp^2}\overline{g}} }{\left|s_\para\right|^{4\D + 2a_{\hp^2}\overline{g}}} + \mco(\overline{g}^2) \ .
\end{aligned}
\end{equation}

The anomalous dimension can be checked by direct computation from \eqref{eq:Zp2} by using \eqref{eq:AnDimDimreg} together with \eqref{eq:FinalBeta}
\begin{equation} \label{eq:BndyAnomDim}
\begin{aligned}
\g_{\hp^2} = -\be \frac{\p}{\p\hg} Z_{\hp^2} = \frac{4\pi(N + 2)A_d^2\overline{g}}{3} + \mco(\hg^2) = \frac{N + 2}{N + 8}\e + \mco(\e^2) \ .
\end{aligned}
\end{equation}

This yields the full scaling dimension of $\hp^2$
\begin{equation}
\begin{aligned}
\D_{\hp^2} = 1 - \frac{6\e}{N + 8} + \mco(\e^2) \ .
\end{aligned}
\end{equation}

\subsubsection{Small Distance Cutoff}

Let us now find the same results, but using a small distance cutoff instead. I.e we $\epsilon-$expand the bulk-bulk correlator \eqref{eq:P2P21} and then study it as one of the points approaches the boundary. In the following we will assume that the coupling is evaluated at the WF fixed point \eqref{eq:FixedPt} with $\hg \propto \epsilon$. In the boundary limit of the bulk-bulk correlator (\ref{eq:Corrs}) we find a logarithmic divergence
\begin{equation*}
\begin{aligned}
\langle\ph^2(x)\ph^2_\Lambda(y_\parallel)\rangle_{1} &= \frac{16\pi N(N + 2)A_3^4\hat{g}}{3}\frac{\log(x_\perp^2) - 2\log(s_\para^2 + x_\perp^2) - 2\log(\Lambda)}{s_\para^2 + x_\perp^2} + \mathcal{O}\left(\epsilon^2, \frac{1}{\Lambda}\right) \; ,
\end{aligned}
\end{equation*}

where we have used the notation introduced in the Section \ref{sec:CutoffReg} with $\ph^2_\Lambda(y_\parallel) \equiv \ph^2(y)|_{y_\perp = \frac{1}{\Lambda}}$. By adding this to the leading order term \eqref{eq:P2free} and resumming we get
\begin{equation}
\begin{aligned}
\langle\ph^2(x)\ph^2_\Lambda(y_\parallel)\rangle &= 8NA_d^2\left(\frac{\La}{\m}\right)^{-a_\ph^2\hg}\frac{\mu^{-a_{\hp^2}\hg} }{\left|x_{\perp}\right|^{-a_{\hp^2}\hg }\left(s_\para^2 + x_\perp^2\right)^{2\D + a_{\hp^2}\hg}} + \mco(\hg^2) \ .
\end{aligned}
\end{equation}

Here $a_{\hp^2}$ is the coefficient \eqref{eq:aPhi2Coeff}. If we now use \eqref{eq:DivergentCorr}\footnote{Note that we have to divide by the normalization constant $8NA_d^2$ from \eqref{eq:P2free} to obtain the correct BOE coefficient.}  this correlator tells us that in the BOE of $\ph^2$ we have the divergent terms
\begin{equation} \label{eq:Phi2BOE}
\begin{aligned}
\ph^2_\Lambda(y_\parallel) &= \left(\frac{\La}{\m}\right)^{-a_\ph^2\hg}\hp^2(y_\parallel) + ... \ .
\end{aligned}
\end{equation}

This yields the same renormalized bulk-boundary correlator as \eqref{eq:BndyIntCorrRes}, but with $\hg$ instead of $\overline{g}$, thus giving us the same boundary anomalous dimensions (\ref{eq:BndyAnomDim}). A practical feature of using a cutoff is that we do not need to do any extra loop integrals in order to find new bulk-boundary correlators of derivative fields. This means that we can proceed to differentiate $\ph^2(y)$ w.r.t. $y_\perp$ and take the boundary limit to every $\ph^2 - \p_\perp^n\hp^2$ correlator without doing a single new loop integral. Let us illustrate how this works in the next section with one derivative, and check that the boundary condition (\ref{eq:ModifiedBC2}) holds.

\subsection{$\ph^4$ correlators} \label{sec:PhiFour}

\subsubsection{Check on boundary condition} \label{sec:BC}

\begin{figure}
	\centering
	\includegraphics[width=0.75\textwidth]{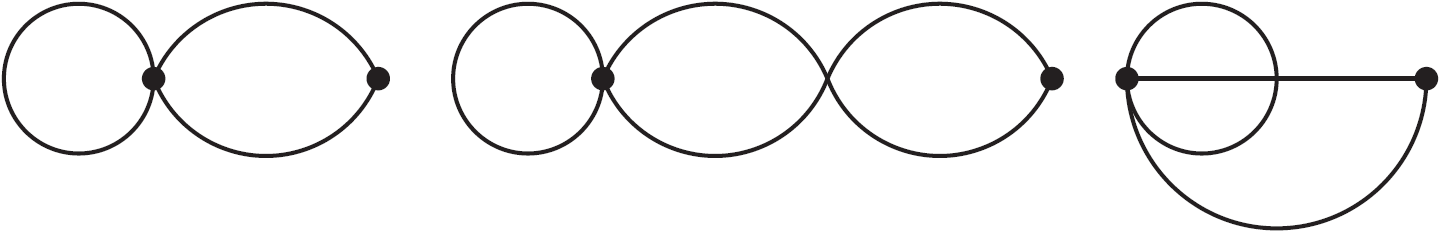}
	\caption{The $\ph^4 - \ph^2$ correlator up to order $\hg$. The full dots represent the composite operator insertions.}
	\label{Fig:P2P4}
\end{figure}

We will verify the modified Neumann boundary condition (\ref{eq:ModifiedBC2}) by studying the boundary limit of $\langle\ph^4(x)\p_\perp\ph^2(y)\rangle \ .$ As discussed before, when dealing with normal derivatives, it is very convenient to regulate divergences in the boundary limit with a small distance cutoff. We will also study the boundary limit of the $\ph^4 - \ph^2$ as well as $\ph^4 - \ph^4$ correlator. The connected part of the $\langle\ph^4(x)\ph^2(y)\rangle$ in the free theory consists of only one connected diagram (see the first diagram on Figure \ref{Fig:P2P4})
\begin{equation} \label{eq:Phi4-Phi2 Corr}
\begin{aligned}
\langle\ph^4(x)\ph^2(y)\rangle &= 2N(N + 2)\langle\ph^2(x)\rangle\langle\ph(x)\ph(y)\rangle^2 \ .
\end{aligned}
\end{equation}

At order $\hg$ there are two connected diagrams depicted in Figure \ref{Fig:P2P4}, which give 
\begin{equation} \label{eq:Phi4-Phi2 Order g}
\begin{aligned}
\langle\ph^4(x)\ph^2(y)\rangle_1 &= -\frac{\hat{g}}{4!}\int d^{d - 1}z_\para \left( 16N(N + 2)^2\langle\ph^2(x)\rangle\langle\ph(x)\hp(z_\para)\rangle^2\langle\hp(z_\para)\ph(y)\rangle^2 + \rig \\
\eq \lef + 64N(N + 2)\langle\ph(x)\ph(y)\rangle\langle\ph(x)\hp(z)\rangle^3\langle\hp(z)\ph(y)\rangle \right) \\
&= -\frac{32N(N + 2)A_d^4\hg}{3} \left( (N + 2)\langle\ph^2(x)\rangle I_{2\D, 2\D}^{d - 1} + 4\langle\ph(x)\ph(y)\rangle I_{3\D, \D}^{d - 1} \right) \ .
\end{aligned}
\end{equation}

Here $I_{\al\be}^{n}$ is the integral in Appendix \ref{app:ph^2Loop}. The boundary limit of this correlator is renormalized if the BOE of $\ph^2$ is on the form \eqref{eq:Phi2BOE}. It yields the resumed renormalized correlator up to order $\hg$
\begin{equation} \label{eq:Phi4-Phi2 BoundLim}
\begin{aligned}
\langle\ph^4(x)\hp^2(y_\parallel)\rangle &= 8N(N + 2)A_d^3 \left( 1 - \frac{32\pi A_d^2\hg}{3} \right)\frac{\mu^{-a_{\hp^2}\hg}}{\left|x_\perp\right|^{2\D - a_{\hp^2}\hg}\left(s_\para^2 + x_\perp^2\right)^{2\D + a_{\hp^2}\hg}} \ .
\end{aligned}
\end{equation}
from which we find the BOE coefficient
\begin{equation} \label{eq:BOEp4p2}
\m^{\phi^4}{}_{\hp^2}= (N + 2)A_d \left( 1 - \frac{32\pi A_d^2\hg}{3} \right) + \mco(\hg^2) = \frac{N + 2}{4\pi} \left( 1 - \frac{12\e}{N + 8} \right) + \mco(\e^2) \; ,
\end{equation}
where we used \eqref{eq:BoundaryCorrelator} and divided by the normalization constant from \eqref{eq:P2free}. 

\quad\quad Let us proceed to calculate the $\ph^4 - \p_\perp\hp^2$ to check the modified boundary condition \eqref{eq:ModifiedBC2}. If we differentiate \eqref{eq:Phi4-Phi2 Corr} w.r.t. $y_\perp$ and then consider the boundary limit we find
\begin{equation}
\begin{aligned}
\lim\limits_{y_\perp\rightarrow 0}\langle\ph^4(x)\p_\perp\ph^2(y)\rangle = 0 \ .
\end{aligned}
\end{equation}

At order $\hg$ we differentiate \eqref{eq:Phi4-Phi2 Order g} w.r.t. $y_\perp$, which gives
\begin{equation} \label{eq:Phi4Phi2}
\begin{aligned}
\langle\ph^4(x)\p_\perp\ph_\La^2(y_\para)\rangle &= \frac{128N(N + 2)A_3^4\hg}{3\left( s^2 + x_\perp^2 \right)^2} + \frac{8N(N + 2)^2A_3^4\hg}{3\left( s^2 + x_\perp^2 \right)x_\perp}\La \ .
\end{aligned}
\end{equation}

\begin{figure}
	\centering
	\includegraphics[width=0.5\textwidth]{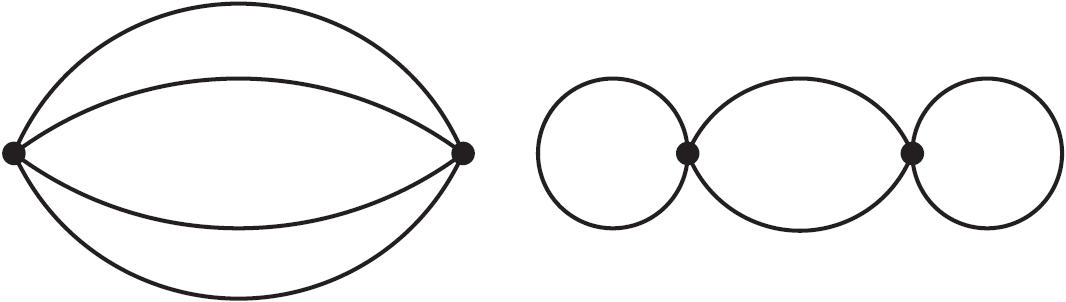}
	\caption{The $\ph^4 - \ph^4$ correlator in the free theory. The full dots represent the respective composite operators.}
	\label{fig:P4P4}
\end{figure}

In order to check the boundary condition, we wish to compare this with the boundary limit of the $\langle\ph^4(x)\ph^4(y)\rangle$. This correlator is computed from the two connected diagrams on figure \ref{fig:P4P4}
\begin{equation*} 
\begin{aligned}
\langle\ph^4(x)\ph^4(y)\rangle &= 8N(N + 2)\langle\ph(x)\ph(y)\rangle^4 + 8N(N + 2)^2\langle\ph^2(x)\rangle\langle\ph(x)\ph(y)\rangle^2\langle\ph^2(y)\rangle \ .
\end{aligned}
\end{equation*}

The boundary limit of this correlator
\begin{equation} \label{eq:FreePhi4Phi4}
\begin{aligned}
\lim\limits_{y_\perp\rightarrow 0}\langle\ph^4(x)\ph^4(y)\rangle &= \frac{128N(N + 2)A_3^4}{\left( s^2 + x_\perp^2 \right)^2} + \frac{8N(N + 2)^2A_3^4}{\left( s^2 + x_\perp^2 \right)x_\perp}\La \ .
\end{aligned}
\end{equation}

\begin{figure} 
	\centering
	\includegraphics[width=0.75\textwidth]{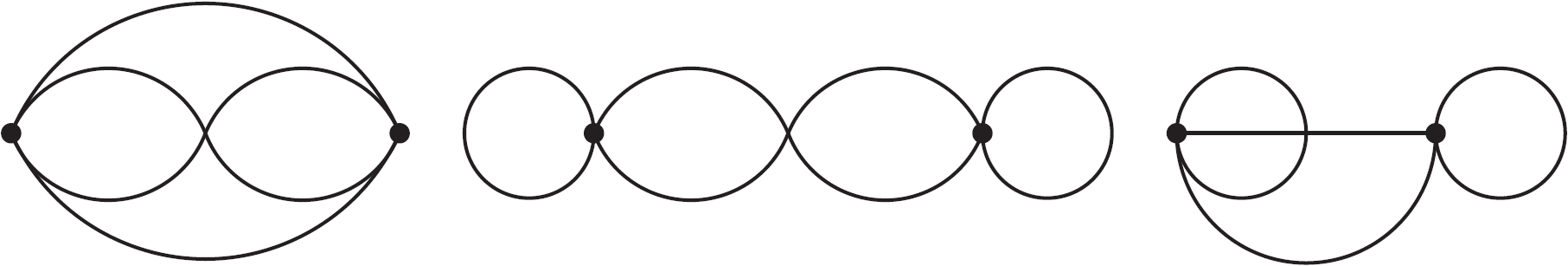}
	\caption{The $\ph^4 - \ph^4-$ correlator at order $\hg \ .$}
	\label{Fig:p4p42loops}
\end{figure}

This is indeed $3^{-1}\hg$ times (\ref{eq:Phi4Phi2}), just as the boundary condition (\ref{eq:ModifiedBC2}) predicted. Having confirmed the boundary condition, let us study the BOE of $\ph^4$. By comparing with \eqref{eq:DivergentCorr} the above correlator tells us that BOE of $\phi^4$ contains $\hp^2$. With the bulk-boundary correlator \eqref{eq:Phi4-Phi2 BoundLim} at hand, we can proceed to study the BOE of $\ph^4$. At order $\hg$ we have the diagrams depicted on Figure \ref{Fig:p4p42loops}
\begin{equation*}
\begin{aligned}
\langle\ph^4(x)\ph^4(y)\rangle_1 &= -\frac{64N(N + 2)A_d^4\hg}{3} \left( 2(N + 2)\langle\ph(x)\ph(y)\rangle \left( \langle\ph^2(x)\rangle I_{\D,3\D}^{d - 1} + \langle\ph^2(y)\rangle I_{3\D,\D}^{d - 1} \right) + \rig \\
\eq \lef + \left( 2(N + 8)\langle\ph(x)\ph(y)\rangle^2 + (N + 2)^2\langle\ph^2(x)\rangle\langle\ph^2(y)\rangle \right) I_{2\D,2\D}^{d - 1} \right) \ .
\end{aligned}
\end{equation*}

We can resum the boundary limit of this correlator added to the free theory one \eqref{eq:FreePhi4Phi4} to get
\begin{equation*} 
\begin{aligned}
\langle\ph^4(x)\ph^4_\La(y_\para)\rangle &= 8A_d^4N(N + 2) \left[ 16 \left( 1 + \frac{4\pi(N + 2)A_d^2\hg}{3} \right) \left( \frac{\La}{\mu} \right)^{-a_{\hp^4}\hg} \frac{\m^{-a_{\hp^4}\hg}}{ x_\perp^{-a_{\hp^4}\hg}\left( s_\para^2 + x_\perp^2 \right)^{4\D + a_{\hp^4}\hg} } + \rig \\
\eq \lef + (N + 2) \left( 1  - \frac{64\pi A_d^2\hg}{3} \right) \frac{\La^{1 - a_{\hp^2}\hg}}{\mu^{- a_{\hp^2}\hg}} \frac{\m^{-a_{\hp^2}\hg}}{ x_\perp^{1 - a_{\hp^2}\hg}\left( s_\para^2 + x_\perp^2 \right)^{2\D + a_{\hp^4}\hg} } \right] \ .
\end{aligned}
\end{equation*}

The $\La^{1 - a_{\hp^2}\hg}$-term reconfirms the  presence of $\hp^2$ in the BOE of $\ph^4$, with the respective coefficient being exactly square of \eqref{eq:BOEp4p2} as expected from \eqref{eq:DivergentCorr}. Additionally by comparing this correlator with \eqref{eq:DivergentCorr} we find that $a_{\hp^2}\hg$ is indeed the anomalous dimension of $\hp^2$, see \eqref{eq:BndyAnomDim}, and that $a_{\hp^4}$ is the anomalous dimension of $\hp^4$
\begin{equation} \label{eq:AnomDim}
\begin{aligned}
\g_{\hp^4} = a_{\hp^4}\hg + \mco(\hg^2) = \frac{8\pi(N + 8)A_d^2\hg}{3} + \mco(\hg^2) = 2\e + \mco(\e^2) \ .
\end{aligned}
\end{equation}

This yields the full scaling dimension of $\hp^4$
\begin{equation}
\begin{aligned}
\D_{\hp^4} = 2 + \mco(\e^2) \ .
\end{aligned}
\end{equation}

In order to get a finite bulk-boundary correlator, we require that the divergent part of the BOE of $\ph^4$ is
\begin{equation*} \hspace{-10px}
\begin{aligned} 
\ph^4_\La(y_\para) &= (N + 2)A_d \left( 1 - \frac{32\pi A_d^2 \hg}{3} \right) \frac{\La^{1 - a_{\hp^2}\hg}}{\mu^{- a_{\hp^2}\hg}} \hp^2(y_\para) + \left( 1 + \frac{4\pi(N + 2)A_d^2\hg}{6} \right)\left( \frac{\La}{\mu} \right)^{-a_{\hp^4}\hg}\hp^4(y_\para) + ... \ . 
\end{aligned}
\end{equation*}
where the dots stand for terms vanishing in the boundary limit as usual. From here we can read off the new BOE coefficients \eqref{eq:BOEp4p2} and
\begin{equation} \label{eq:BOEp4p4}
\m^{\phi^4}{}_{\hp^4}= 1 + \frac{4\pi(N + 2)A_d^2\hg}{6} + \mathcal{O}(\hg^2) = 1 + \frac{(N + 2)\e}{2(N + 8)} + \mathcal{O}(\e^2) \; ,
\end{equation}

This yields the bulk-boundary correlator
\begin{equation}
\begin{aligned}
\langle\ph^4(x)\hp^4(y_\para)\rangle &= 128A_d^4N(N + 2) \left( 1 + \frac{4\pi(N + 2)A_d^2\hg}{3} \right)  \frac{\m^{-a_{\hp^4}\hg}}{ x_\perp^{-a_{\hp^4}\hg}\left( s_\para^2 + x_\perp^2 \right)^{4\D + a_{\hp^4}\hg} } \ .
\end{aligned}
\end{equation}

\subsubsection{Consistency with dimensional regularization} \label{sec:DimregConsistency}

From \eqref{eq:CouplingCond} we expect that the respective renormalization factor is related to the coupling renormalization so here we would like to verify this. We start by writing
\begin{equation} \label{eq:Zp4def1}
\lim_{x_\perp \to 0} Z_{\hp^4}\ph^4 = \hp^4 \; ,
\end{equation}
where $Z_{\hat{\phi}^4}$ contains poles in $\epsilon$. This divergent quantity can be found by using \eqref{eq:CouplingCond} and the data computed in Section \ref{sec:BetaDeriv} since 
\begin{equation} \label{eq:dg0dg}
Z_{\hat{\phi}^4}=\mu^{d - 3}\frac{\partial}{\partial \hat{g}} \hat{g}_0 =  \mu^{d - 3} \frac{(d - 3)\hat{g}_0}{\beta}= Z_{\hat{g}} \left( 1 + \hg\frac{\partial}{\partial g} \ln Z_{\hat{g}} \right) \; ,
\end{equation}

where we have used the definition of $\hat{g}_0$ given in \eqref{eq:g0Def}. The expression in \eqref{eq:g0Def} is evaluated by substituting the $Z_{\hat{g}}$ we found in Section \ref{sec:BetaDeriv} so 
\begin{equation} \label{eq:dg0dg1}
Z_{\hat{\phi}^4} = 1+ \frac{2 a_{\hg}}{\epsilon}+ \mathcal{O}(\hg^2) \; ,
\end{equation}
where $a_{\hg}$ was given in \eqref{eq:Z1Coupling}.

\quad\quad On the other hand $Z_{\hat{\phi}^4}$ can be found from the $\epsilon$ pole in the $\ph^4-\hp^4$ correlator in dimreg. Since there are no power divergences in the dimreg limit, we only need to consider the diagram without self-closing loops on Figure \ref{Fig:p4p42loops}.\footnote{The boundary limit of $\vev{\phi^2}$ is zero in dimreg, which can be seen as fine tuning of power divergent boundary mass term.} We can renormalize the boundary limit of $\ph^4$ using dimreg in the same way as $\ph^2$ in Section \ref{sec:dimregPhi2}. This reproduces the finite bulk-boundary correlator $\langle\ph^4(x)\hp^4(y)\rangle$ from the Section \ref{sec:PhiFour},\footnote{Where we have replaced the coupling constant with the MS bar coupling (\ref{eq:MSBarCoupling}).} with the following $Z$-factor
\begin{equation}
\begin{aligned}
 Z_{\hp^4} = 1 + a_{\hp^4}\frac{\overline{g}}{\e} \ . 
\end{aligned}
\end{equation}

Here $a_{\hp^4}$ is the constant (\ref{eq:AnomDim}). This gives the same anomalous dimension for $\hp^4$ as in the previous Section using \eqref{eq:AnDimDimreg}
\begin{equation}
\gamma_{\hp^4}= - \beta \frac{\partial}{\partial g}  Z_{\hp^4} = a_{\hp^4}\overline{g} + \mathcal{O}(\overline{g}^2) \; .
\end{equation}
 The leading divergent piece of the $Z_{\hp^4}$ reads
\begin{equation}
\begin{aligned}
a_{\hp^4} = \frac{8\pi(N + 8)A_3^2}{3} = \frac{N + 8}{6\pi} = 2a_{\hg} 
\end{aligned}
\end{equation}
as predicted by \eqref{eq:dg0dg1}.

\section{Stress-energy tensor} \label{Sec:SETensor}

The bulk SE tensor and its correlators can be defined via coupling the theory with action $S$ to a metric $g_{\mu \nu}$ and defining
\begin{equation}
T^{\mu \nu}(x)= - \left. 2 \frac{1}{\sqrt{g}}\frac{\delta S}{\delta g_{\mu \nu}(x)} \right|_{g_{\mu \nu}= \delta_{\mu \nu}} \; .
\end{equation}

Or alternatively we can perform a variation of the action 
\begin{equation}
\delta S = - \frac{1}{2} \int d^d x \sqrt{g} \delta g_{\mu \nu} T^{\mu \nu}(x) \; .
\end{equation}

For conformally coupled scalar field we have
\begin{equation}
S= \frac{1}{2}\int d^d x \sqrt{g} ( (\partial \phi)^2 + \xi R \phi^2) \; , \quad \x = \frac{1}{4}\frac{d - 2}{d - 1} \ .
\end{equation}

Using the standard variational principle and integration by parts we find
\begin{equation} \label{eq:BulkScTensor}
T^{\mu \nu}= \partial^\mu \phi \partial^\nu \phi - \frac{1}{2} \delta^{\mu \nu} (\partial \phi)^2 - \xi (\partial^\mu \partial^\nu - \delta^{\mu \nu} \Box) \phi^2 \; ,
\end{equation}

The presence of $\xi-$term guarantees vanishing of the stress-tensor trace
\begin{equation}
T_\mu^\mu=\frac{(d-2)}{2} \phi \Box \phi \; ,
\end{equation}

which is proportional the equation of motion.

\subsection{Renormalization of boundary stress-energy tensor} \label{sec:TenRen}

In this section we will determine renormalized boundary SE tensor $\hat{\tau}^{ab}$ in dimreg. The bulk SE tensor (with improvement) is precisely \eqref{eq:BulkScTensor}. Let us now find the l.h.s. of \eqref{eq:MomentumCond} for this model. The relevant bulk operator reads
\begin{equation} \label{eq:BoundSE}
T^{a \perp}= \partial^a \phi \partial^\perp \phi - \x\partial^a \partial^\perp \phi^2 \; .
\end{equation}

Taking the boundary limit of the above expression and using \eqref{eq:ModifiedBC2} leads to
\begin{equation}
\lim_{x_\perp \to 0} T^{a \perp} = \left(1 - 8\x \right) \frac{\hg_0}{4!} \partial^a O_4 = -\frac{d - 3}{d - 1}\frac{\hg_0}{4!} \partial^a O_4 \; ,
\end{equation}

where 
\begin{equation}
O_4 = \lim_{x_\perp \to 0} \ph^4 \; .
\end{equation}

By comparing this with \eqref{eq:MomentumCond} we deduce that
\begin{equation} \label{eq:ModelBTensor}
\hat{\tau}^{ ab}= -\frac{d - 3}{d - 1}\frac{\hg_0}{4!}\delta^{ab}O_4 \; ,
\end{equation}

which appears to be divergent. In Section \ref{sec:DimregConsistency} we observed that the near-boundary divergences of $O_4$ in dimreg follow from the RG equation of the quartic coupling. Indeed from \eqref{eq:Zp4def1} and \eqref{eq:dg0dg} we get that 
\begin{equation}
O_4= Z_{\hat{\phi}^4}^{-1} \hp^4= \left((d-3) \frac{\hat{g}_0}{4!} \right)^{-1} \hp^4 \; .
\end{equation}

This can be inserted back into \eqref{eq:ModelBTensor} to give
\begin{equation} \label{eq:tabRen}
\hat{\tau}^{ ab}= \frac{\beta \hat{\phi}^4 \delta^{ab}}{d - 1}  \; 
\end{equation}

and 
\begin{equation} \label{eq:TnaModel}
\lim_{x_\perp \to 0} T^{a \perp}= \frac{\beta \partial^a \hat{\phi}^4}{d - 1} \; ,
\end{equation}
which is exactly of the form \eqref{eq:TnaBOE1}.
Few remarks are in order. First, the r.h.s. of \eqref{eq:TnaModel} is manifestly finite at all orders in perturbation theory and so is \eqref{eq:tabRen}.\footnote{Had we used a cutoff regularization we would have to take into account linearly divergent contribution to $\lim_{x_\perp \to 0} T^{a \perp}$ coming from the contribution of $\partial^a\phi^2$ to the BOE. This can be cancelled by fine tuning the boundary mass term \eqref{eq:PowerDivPP}.} In particular this means that $\hat{\tau}^{ab}$ doesn't acquire any anomalous dimension. Secondly the conformal boundary condition \eqref{eq:CardyCon} is satisfied for the fixed point \eqref{eq:Be0} studied in this paper, which proves the conformal invariance of the model.
Finally let us note that for $\epsilon>0$ where the theory has an interacting fixed point \eqref{eq:FixedPt} we can solve the RG equation for the beta function \eqref{eq:FinalBeta} in the IR limit described in \eqref{eq:IRlimit}
\begin{equation} \label{eq:BetaRGsoln}
\beta \approx \left(\frac{\mu}{\Lambda}\right)^{\epsilon}  \; ,
\end{equation}
where $\Lambda= \frac{1}{x_\perp}$ near-boundary cutoff. By substituting \eqref{eq:BetaRGsoln} in \eqref{eq:TnaModel} we get
\begin{equation}
 T^{a \perp} \stackrel{x_\perp \to 0}{\sim} (x_\perp \mu)^{\epsilon} \partial^a \hp^4 \; ,
\end{equation}
which is exactly of the form \eqref{eq:TnaBOE1} and corresponds to the fact the deformation $\hat{\phi}^4$ becomes an irrelevant operator close to the IR fixed point.

\subsection{Consistency of boudary stress-energy tensor with bulk correlators} \label{sec:SE}
In the previous section we derived the all order relation \eqref{eq:TnaModel} indirectly by using RG equations. In this section we would like to check the vanishing of l.h.s. of \eqref{eq:TnaModel} at the WF fixed point \eqref{eq:FixedPt} in a specific correlator. Namely we would like to evaluate  $\langle\ph^4(x)T^{a\perp}(y)\rangle$ in the boundary limit. To actually check the vanishing of l.h.s. with the full beta function \eqref{eq:FinalBeta} we would need to compute three-loop (or $\mathcal{O}(\hg^2)$) diagrams. Instead we will just check that it vanishes at leading order in $\e-$expansion in a manner consistent with \eqref{eq:TnaModel}. By substituting the expression for beta function found in \eqref{eq:FinalBeta} in \eqref{eq:TnaModel} we get
\begin{equation} \label{eq:TnaLeading}
\lim_{x_\perp \to 0} T^{a \perp}= -\frac{ \epsilon \hg}{2} \partial^a \hp^4 =  0+\mathcal{O}(\e^2) \; .
\end{equation}
Based on the above our goal to verify that the two-point $\lim_{y_\perp \to 0}$ of $\langle\ph^4(x) T^{a\perp}(y_\perp)\rangle$ vanishes to the leading order in $\epsilon$. 
As we can see from \eqref{eq:TnaLeading}, we expect a non-zero contribution at the order $\e \hat{g} \ ,$ proportional to $\langle\ph^4\p_\para^a\hp^4\rangle \ ,$ which we found in Section \ref{sec:BC}
\begin{equation} \label{eq:FreePhi^4DerPhi^4Corr}
\begin{aligned}
&\langle\ph^4(x)\p_\para^a\hp^4(y_\para)\rangle = \frac{1028\D N(N + 2)A_d^4s^a}{\left(s_\para^2 + x_\perp^2\right)^{4\D + 1}} + \mco(\hg) = \frac{512N(N + 2)A_3^4s^a}{x_\perp^6} + \mco(\e, s_\para^2, \hg) \ .
\end{aligned}
\end{equation}

This means that we expect from (\ref{eq:ModelBTensor})
\begin{equation} \label{eq:CardyCond}
\begin{aligned}
&\langle\ph^4(x)\hat{T}_{\para\perp}^a(y_\para)\rangle = \frac{32N(N + 2)A_3^4\e\hat{g}s^a}{3x_\perp^6} + \mco(\e^2, \hat{g}^2, s_\para^2) \ .
\end{aligned}
\end{equation}

We expand the correlator in a small parallel distance expansion in order to solve the loop integrals in $\langle\ph^4(x)\hat{T}^{a\perp}(y_\para)\rangle$. Let us first check that this correlator is zero in the free theory. Using \eqref{eq:BoundSE} we find
\begin{equation*}
\begin{aligned}
\langle\ph^4(x)T^{a\perp}(y)\rangle &=4N(N + 2)\langle\ph^2(x)\rangle \left( \p^a_{y_\para}\langle\ph(x)\ph(y)\rangle\p_{y_\perp}\langle\ph(x)\ph(y)\rangle - \x\p_{y_\perp}\p^a_{y_\para}\langle\ph(x)\ph(y)\rangle^2 \right)
\end{aligned}
\end{equation*}

Due to the boundary condition \eqref{eq:ModifiedBC}, the boundary limit of above correlator is trivial
\begin{equation}
\begin{aligned}
\langle\ph^4(x)\hat{T}^{a\perp}(y)\rangle &= 0 \ .
\end{aligned}
\end{equation}

Let us now proceed to order $\hat{g} \ .$ Only the connected diagrams will yield non-trivial boundary limits after renormalization
\begin{equation*}
\begin{aligned}
\langle\ph^4(x)T^{a\perp}(y)\rangle_1 &= -\frac{\hat{g}}{4!}\int d^{d - 1}z_\para \left( \langle\ph^4(x)\hp^4(z_\para)\p^a_\para\ph\p_\perp\ph(y)\rangle - \xi\langle\ph^4(x)\hp^4(z_\para)\p^a_\para\p_\perp\ph^2(y)\rangle \right) \ .
\end{aligned}
\end{equation*}

Details on this loop calculation are in Appendix \ref{App:SE}. This correlator has the boundary limit
\begin{equation} \label{eq:BoundLim}
\begin{aligned}
\lim_{y_\perp \to 0}\langle\ph^4(x)\hat{T}^{a\perp}\rangle &= \frac{128\pi N(N + 2)A_3^5\e\hat{g}}{3}\frac{s_\para^a}{x_\perp^6} + \mco(\e^2, s_\para^2, \hg^2) \ .
\end{aligned}
\end{equation}
Which agrees with (\ref{eq:CardyCond}) due to (\ref{eq:Const3Dim}). 

\subsection{Displacement Operator} \label{sec:Disp}

The displacement operator exist in any bCFT, and should be protected. It is in the BOE of the perpendicular components of the SE tensor at the boundary, which we find from \eqref{eq:BulkScTensor} using the bulk equations of motion \eqref{eq:ModifiedBC} 
\begin{equation}
\begin{aligned}
T^{\perp\perp} &= (\partial_\perp \phi)^2 - \frac{\p_\perp^2\ph^2}{4} - \left( \frac{1}{4} - \x \right) \p_\para^2\ph^2 = (\partial_\perp \phi)^2 - \frac{1}{4} \left( \p_\perp^2\ph^2 + \frac{\p_\para^2\ph^2}{d - 1} \right) \ .
\end{aligned}
\end{equation}
In order to define correlators of the displacement operator, one needs to study the boundary limit of above operator. We will do this in the example of $\ph^2 - T^{\perp\perp}$ correlator, which is found by differentiating (\ref{eq:Corrs}). Using the boundary cutoff we find the leading nonvanishing part of the correlator 
\begin{equation} 
\begin{aligned}
\langle\ph^2(x)\hat{D}(y_\para)\rangle &= \lim\limits_{y_\perp\rightarrow 0}\langle\ph^2(x)T^{\perp\perp}(y)\rangle = -\frac{4N(d - 2)^2A_d^2x_\perp^2}{\left( s_\para^2 + x_\perp^2 \right)^d} + \mco(\hg^2) \ ,
\end{aligned}
\end{equation}

which corresponds to an operator of dimension $d$ by comparison with \eqref{eq:BoundaryCorrelator} (assuming that $\Delta_{\phi^2}=(d-2)$). With the boundary condition \eqref{eq:DisplacementDef} in mind we should identify this term as the contribution of the displacement operator $\hat{D}$.
The operator $\p_\para^2\hp^2$ is a descendant of $\hp^2$, and its contribution comes from the differential operator in the BOE, see (\ref{eq:BOE}). From above correlator we see that $\hat{D}$ and its BOE coefficient are indeed protected at this order (the operator $\hat{D}$ on the r.h.s. defined through the variational rule \eqref{eq:DisplacementDef1} is physical so we do not expect it to acquire any divergences or anomalous dimension).
Another interesting object we can compute is the two-point function of the displacement operator, whose coefficient is related to a $B-$type conformal anomaly (a central charge) at the boundary in $d=3$ (see \cite{Bianchi:2015liz},   \cite{Herzog:2017kkj} for more details). In our case this two-point function reads 
\begin{equation} \label{eq:DDcorrResult}
\begin{aligned}
\langle\hat{D}(x_\para)\hat{D}(y_\para)\rangle = \frac{2N(d - 2)^2A_d^2}{\left|s_\para\right|^{2d}} + \mathcal{O}(\hg^2) \ ,
\end{aligned}
\end{equation}

from which we see that the central charge does not receive any corrections at $\mathcal{O}(\hg)$. For completeness let us write the value of $\vev{\hat{D} \hat{D}}$ for different dimensions
\begin{equation} \label{eq:TnnTnnDimension}
\begin{aligned}
d = 2 \quad\Rightarrow\quad \langle\hat{D}(x_\para)\hat{D}(y_\para)\rangle &= \frac{N}{2\pi^2s_\para^4} + \mathcal{O}(\hg^2) \ , \\
d = 3 \quad\Rightarrow\quad \langle\hat{D}(x_\para)\hat{D}(y_\para)\rangle &= \frac{N}{8\pi^2s_\para^6} + \mathcal{O}(\hg^2) \ , \\
d = 4 \quad\Rightarrow\quad \langle\hat{D}(x_\para)\hat{D}(y_\para)\rangle &= \frac{N}{2\pi^4s_\para^8} + \mathcal{O}(\hg^2) \ .
\end{aligned}
\end{equation}
Note that the $d=3$ expression agrees with the free-field computation of \cite{1709.07431}.

\section{Discussion and outlook} 

\subsection{Physical significance of the model} \label{sec:Physical}

Let us conclude by commenting on possible interpretation of the model presented here as we analytically continue in $\epsilon$. In analogy with the bulk $\epsilon-$expansion
we might wonder whether the fixed point we analysed is not in a same universality class as some bCFT in higher/lower dimension. 

\quad\quad The $\epsilon=1$ limit should correspond to a $c=1$ free bosonic bCFT with a boundary potential. A complete classification for such theories was proposed in \cite{Gaberdiel:2001zq}. Presently it is not  clear to us how the model fits into this classification, but let us offer a qualitative argument of what kind of bCFT one might expect. To actually understand the nature of the RG flow described in Section \ref{sec:BetaDeriv} at $\epsilon=1$ one needs to interpret the operator $\hp^4$ (cf. Section \ref{sec:PhiFour}) that is driving the RG flow. In particular this operator will have a bare dimension $0$ in $d=2$ (and so will all $\phi^n$s). This means that it can mix with all the other $\phi^{2n}$s under renormalization or more precisely in the language of Section \ref{sec:Divergences}
\begin{equation} \label{eq:ZnP2n}
\lim_{x_\perp \to 0} \sum_{n} Z_{4}^n \phi^{2n} = \hp^4 \; ,
\end{equation}
where $Z_{4}^n$ are the renormalization constants. In particular the l.h.s. of \eqref{eq:ZnP2n} can actually come from Taylor expansion of a function $f(\phi^{2})$. For such general boundary potentials there might exist non-trivial interacting bCFTs (see for example \cite{Callan:1994ub}). Furthermore we see that the operator becomes irrelevant with positive anomalous dimension $\approx 2$ from \eqref{eq:AnomDim},\footnote{Although this value might change drastically with the inclusion of higher order corrections in $\epsilon$} whereas the dimension of $\hp^2$ is slightly below $1$ (at least for large enough $N$), which is perfectly consistent with unitarity bounds. Another consistency check is our result for $\vev{\hat{D} \hat{D}}$ at $d=2$ (see \eqref{eq:DDcorrResult}), which does not receive any corrections up to $\mathcal{O}(\hg)$. This is consistent with the fact that in $d=2$ the correlator $\vev{\hat{D} \hat{D}}$ is determined from the bulk central charge \cite{9302068}, which is just the free scalar one in our case. 
In any case to make more solid conclusions it would be beneficial to perform the analysis at higher orders in $\epsilon$.

\quad\quad On the other hand as already mentioned in Section \ref{sec:BetaDeriv} the UV fixed point at $\epsilon=-1$ involves a quartic potential with negative sign and therefore the model in $d=4$ should be unstable. A similar issue appears when one studies the bulk $O(N)$ model with quartic interaction in $d=4+\epsilon$ \cite{Fei:2014yja}. In this work a possible resolution was offered by interpreting this CFT as an IR fixed point of some other, well-defined theory instead. One might imagine that a similar mechanism could apply here.

\quad\quad At last let us comment on $\epsilon=0$. Here it would be interesting to study the model in the presence of \textit{bulk} $\phi^4$. One might hope that the contribution of bulk coupling stabilizes the UV behaviour of $\hg$ leading to a non-trivial UV fixed point. In this case the bulk theory gives rise to a chain of dualities \cite{Karch:2016sxi, Seiberg:2016gmd} which might carry over to non-trivial dualities between the boundary states.

\subsection{Outlook} \label{sec:Outlook}

In this paper we explored renormalization of the boundary limits of bulk fields. This is very general, and thus has a lot of applications, e.g. is should work in the same way for i(nterface)CFTs, where a bulk theory is on each side of the boundary. It could be that one needs to generalize these methods in some specific models, e.g. those with degrees of freedom that are restricted to only live on the boundary such as the ones studied in \cite{1707.06224}. 
It would also be interesting to understand the dependence of boundary couplings upon the bulk ones and the interplay between their renormalizations. 
While the model we studied in Section \ref{sec:BoundaryExample} is non-supersymmetric there is nothing preventing us to supersymmetrize the present analysis. Inclusion of fermions is indeed very tempting as then one might want to study what happens to the conserved bulk currents near the boundary and whether an analysis similar to the one we did in Section \ref{sec:SEtensorGeneral} applies. 

\quad\quad Another generalization of the methods in this paper would be to consider CFTs with defects of codimension higher than two, and study how one can renormalize and classify defect fields. An example of such a model where a bulk-bulk correlator is known is the 3D Ising twist defect \cite{Billo:2013jda, 1310.5078, Liendo:2019jpu} or its generalization to $O(N)$ twist \cite{1706.02414}. In this case one could consider taking the norm of the normal coordinates to zero. Though this limit will most likely remember in what direction the bulk field comes from, which means Lorentz invariance is at a risk of breaking in this limit. It seems like it would be useful to first identify what kind of defect fields one can get before considering the defect limit. In a bCFT this is not a problem, since bulk fields can only approach the boundary from one direction. 

\quad\quad The study of two-point functions in the presence of running boundary couplings can be related to the flow of certain conformal anomalies \cite{Jensen:2015swa}. It would be worthwhile to understand whether one can apply the RG methods used in this paper to prove a gradient flow formula along the lines of \cite{Osborn:1991gm} for boundary RG flows.  

\quad\quad There are also lots of other aspects of bCFTs one could investigate, e.g. the role of image symmetry. Does this symmetry restrict the form of two-point correlators? What does it mean to break this symmetry? There is also the mixing problem in the model we discussed in section \ref{sec:BulkModel} that would be interesting to resolve. One way to solve the issue would be to extract the data on these operators by studying suitable correlators in the boundary limit. For example one might obtain their anomalous dimensions from logarithmic divergences of their two-point functions at $\mathcal{O}(\epsilon)$ (cf. \eqref{eq:DivergentCorrPert}). This requires a lot of correlators, but it could be that they all follow a certain behaviour. 

\subsection*{Acknowledgements}

The authors are thankful to Agnese Bissi, Guido Festuccia, Marco Meineri, Andy O'Bannon for commenting on the manuscript and Pierre Vanhove for discussions on the sunrise diagrams. Additionally the authors thank  Emtinan Elkhidir, Davide Gaiotto, Tobias Hansen, Edoardo Lauria, Joe Minahan and Maxim Zabzine  for useful conversations. 
VP is supported by the ERC STG grant 639220 (curvedsusy) and AS is supported by Knut and Alice Wallenberg Foundation KAW 2016.0129

\newpage

\appendix

\section{General form of a bulk-boundary correlator} \label{app:BulkBdyCorr}

We can find the expression of the bulk-boundary correlator from the BOE. The form of the BOE is known from literature e.g. \cite{1601.02883}, and follows from the symmetries preserved by the boundary.\footnote{See Appendix B in \cite{1601.02883}, and set the spin $s$ to zero and the dimension of the defect $p$ to $d - 1 \ .$ We chose this reference since it writes out the differential operator in the BOE as well.} For a scalar in the bulk
\begin{equation} \label{eq:BOE}
\begin{aligned}
O(x) &= \sum_{\hat{O}'}\frac{\m^O{}_{\hat{O}'}}{\left|x_\perp\right|^{\D - \hD'}}B_{\hD'}(x_\perp^2, \p_\para^2)\hat{O}'(x_\para) \ , \\
B_\hD(a, b) &= \sum_{m\geq 0}\frac{(-1)^m}{m!\left(\hD + 1 - \frac{d - 1}{2}\right)_m}\left(\frac{x_\perp}{2}\right)^{2m}\p_{x_\para}^{2m} \ .
\end{aligned}
\end{equation}

Here $(x)_n$ is the Pochhammer symbol. In the free theory, the BOE is a Laurent expansion. It yields
\begin{equation} \label{eq:Bulk-BoundaryCorrfromBOE}
\begin{aligned}
\langle O(x)\hat{O}(y)\rangle &= \sum_{\hat{O}'}\frac{\m^O{}_{\hat{O}'}}{x_\perp^{\D - \hD'}}\sum_{m\geq 0}\frac{(-1)^m}{m!\left(\hD' + 1 - \frac{d - 1}{2}\right)_m}\left(\frac{x_\perp}{2}\right)^{2m}\de^{\hat{O}'}{}_{\hat{O}}\p_{x_\para}^{2m}\left|s_\para\right|^{-2\hD} \ .
\end{aligned}
\end{equation}

One finds that
\begin{equation}
\begin{aligned}
\p_{x_\para}^{2}\left(s_\para^2\right)^{-\hD} = \frac{4\hD\left(\hD + 1 - \frac{d - 1}{2}\right)}{\left|s_\para\right|^{2\left(\hD + 1\right)}} \ .
\end{aligned}
\end{equation}

Applying this derivative formula $m$ times
\begin{equation}
\begin{aligned}
\p_{x_\para}^{2m}\left(s_\para^2\right)^{-\hD} &= \frac{4^m(\hD)_m\left(\hD + 1 - \frac{d - 1}{2}\right)_m}{\left|s_\para\right|^{2\left(\hD + m\right)}} \ .
\end{aligned}
\end{equation}

We can now simplify (\ref{eq:Bulk-BoundaryCorrfromBOE})
\begin{equation}
\begin{aligned}
\langle O(x)\hat{O}(y)\rangle &= \frac{\m^O{}_{\hat{O}}}{x_\perp^{\D - \hD}}\sum_{m\geq 0}\frac{(-1)^m(\hD)_m}{m!}\frac{x_\perp^{2m}}{\left|s_\para\right|^{2\left(\hD + m\right)}}  \ .
\end{aligned}
\end{equation}

The sum can be recognized as the Taylor expansion of $\left(s_\para^2 + x_\perp^2\right)^{-\hD}$ with respect to $x_\perp$
\begin{equation}
\begin{aligned}
\frac{1}{\left(s_\para^2 + x_\perp^2\right)^{\hD}} &= \sum_{m\geq 0}\frac{(-1)^m(\hD)_m}{m!}\frac{x_\perp^{2m}}{\left|s_\para\right|^{2\left(\hD + m\right)}}  \ .
\end{aligned}
\end{equation}

And thus we find the bulk-boundary correlator (\ref{eq:BoundaryCorrelator}).

\section{Four-point correlator} \label{App:4PtCorr}

The amputated four-point correlator is\footnote{Let us point out that the minus sign in the vertex factor comes from the $e^{-S}$ inside the path integral.}
\begin{equation*}
\begin{aligned}
G_4^{ijkl}(\{p_i\}) &= \langle\hp^i(p_1)\hp^j(p_2)\hp^k(p_3)\hp^l(p_4)\rangle \\ 
&= \oo_d^2D^{ijkl} - \frac{\hat{g}_0}{4!}\oo_d^48D^{ijkl} + \left(-\frac{\hat{g}_0}{4!}\right)^2\oo_d^632 \left( E^{ijkl}I_{12} + E^{ikjl}I_{13} + E^{iljk}I_{14} \right) \ .
\end{aligned}
\end{equation*}

Its $O(N)$-tensor structures are
\begin{equation}
\begin{aligned}
D^{ijkl} &= \de^{ij}\de^{kl} + \de^{ik}\de^{jl} + \de^{il}\de^{jk} \ , \\
E^{ijkl} &= (N + 2)\de^{ij}\de^{kl} + 2D^{ijkl} \ .
\end{aligned}
\end{equation}

and the loop integrals is
\begin{equation}
\begin{aligned}
I_{ij} &= \int \frac{d^{d - 1}k}{(2\pi)^{d - 1}}\frac{1}{|k|^{d - 1 - 2\hD}|k + p_i + p_j|^{d - 1 - 2\hD}} \ .
\end{aligned}
\end{equation}

With Feynman parametrization one finds that this integral is given by a Euler-Beta function, which for simplicity we denote $B(x, y) \equiv B_{x, y}$
\begin{equation}
\begin{aligned}
\int_{\mathbb{R}^n}\frac{d^nz}{|z|^{2a}|z + x|^{2b}} &= \frac{\G_{a + b - n/2}}{\G_a\G_b}B_{n/2 - a, n/2 - b}\frac{\pi^{n/2}}{|x|^{2(a + b) - n}} \ .
\end{aligned}
\end{equation}

Which yields
\begin{equation}
\begin{aligned}
I_{ij} &= \frac{B_{\hD, \hD}}{(4\pi)^{(d - 1)/2}}\frac{\G_{(d - 1)/2 - 2\hD}}{\G_{(d - 1)/2  - \hD}^2}\frac{1}{|p_i + p_j|^{d - 1 - 4\hD}} \ .
\end{aligned}
\end{equation}

Let us now specify to $3 - \e$ dimensions
\begin{equation}
\begin{aligned}
\hD = \frac{1 - \e}{2} \quad\Rightarrow\quad I_{ij} = \frac{1}{2\pi}\left( \frac{1}{\e} - \frac{\g_E - \log\left(64\pi\right) + 2\log|p_i + p_j|}{2} + \mco(\e^2) \right) \ .
\end{aligned}
\end{equation}

We can absorb the non-divergent constants in a dimensionless $\overline{MS}$ coupling constant
\begin{equation}
\begin{aligned}
\hat{g}_0^2I^{ij} &= \frac{\bar{g}_0^2}{2\pi}\left( \frac{1}{\e} - \log\left|\frac{p_1 + p_2}{\m}\right| + \mco(\e^2) \right) \ , \quad \hat{g}_0 = \left(\frac{e^{\g_E}}{64\pi}\right)^{\e/4}\m^{\e}\bar{g}_0 + \mco(\e^2) \ .
\end{aligned}
\end{equation}

Absorb the divergence in a renormalized coupling constant (see \eqref{eq:Renorm})
\begin{equation}
\begin{aligned}
\bar{g}_0 &= Z_{\hat{g}}\hat{g} \ . 
\end{aligned}
\end{equation}

Now demand that the correlator is finite
\begin{equation*} \hspace{-30px}
\begin{aligned}
G_4^{ijkl}(\{p_i\}) &= \oo_d^2 \left[ D^{ijkl}  - \frac{\oo_d^2\hat{g}}{3}D^{ijkl} + \frac{\oo_d^2\hat{g}^2}{36\pi} \left( \frac{(N + 8)\oo_d^2 - 12\pi a_{\hg}}{\e}D^{ijkl} - \oo_d^2 \left( F^{ijkl}_{12} + F^{ikjl}_{13} + F^{iljk}_{14} \right)\right)\right] \ ,
\end{aligned}
\end{equation*}
\begin{equation}
\begin{aligned}
F^{ijkl}_{ab} &= E^{ijkl}\log\left|\frac{p_a + p_b}{\m}\right| \ .
\end{aligned}
\end{equation}

This correlator is finite if \eqref{eq:Z1Coupling} holds, which gives us the renormalized correlator (\ref{eq:RenormCorr}).

\section{Master integral} \label{app:ph^2Loop}

The loop integral is
\begin{equation}
\begin{aligned}
I_{\al, \be}^n &\equiv \int_{\mathbb{R}^{n}}d^{n}z_\para\frac{1}{\left[\left(x_\para - z_\para\right)^2 + x_\perp^2\right]^{\al}}\frac{1}{\left[\left(z_\para - y_\para\right)^2 + y_\perp^2\right]^{\be}} \ .
\end{aligned}
\end{equation}

In order to solve the integral $I_{\al\be}$ we first shift the integration variable as $z_\para^a \rightarrow z_\para^a + x_\para^a$ followed by a Feynman parametrization. We can then complete a square in the denominator and make yet another shift $z_\para^a \rightarrow z_\para^a - us_\para^a \ ,$ where $u$ is the Feynman parameter to bring it on the form
\begin{equation*}
\begin{aligned}
I_{\al, \be}^n &= \frac{\G_{\al + \be}}{\G_\al\G_\be}\int_{0}^{1}du(1 - u)^{\al - 1}u^{\be - 1}\int_{\mathbb{R}^n}\frac{d^{n}z_\para}{\left( z_\para^2 + u(1 - u)s_\para^2 + ux_\perp^2 + (1 - u)y_\perp^2 \right)^{\al + \be}} \ .
\end{aligned}
\end{equation*}

The integral over $z_\para$ can be done using Julian-Schwinger parametrization. It is
\begin{equation} \label{eq:MyFavInt}
\begin{aligned}
\int_{\mathbb{R}^d}\frac{d^dz}{\left(z^2 + \D\right)^n} &= \frac{\G_{n - d/2}}{\G_n}\frac{\pi^{d/2}}{\D^{n - d/2}} \ .
\end{aligned}
\end{equation}

This brings the loop integral to the form
\begin{equation} \label{eq:LoopInt}
\begin{aligned}
I_{\al, \be}^n &= \eta^n_{\al, \be}J^n_{\al, \be} \ , \\ 
\eta^n_{\al, \be} &\equiv \pi^{n/2}\frac{\G_{\al + \be - n/2}}{\G_\al\G_\be} \ , \\ 
J^n_{\al, \be} &\equiv \int_{0}^{1}du\frac{u^{\al - 1}(1 - u)^{\be - 1}}{\left( u(1 - u)s_\para^2 + ux_\perp^2 + (1 - u)y_\perp^2 \right)^{\al + \be - n/2}} \ .
\end{aligned}
\end{equation}

The integral $J^n_{\al, \be}$ has a compact closed expression in terms of a hypergeometric functions if any of $s_\para^2 \ , x_\perp^2$ or $y_\perp^2$ is zero. If both $x_\perp$ and $y_\perp$ are zero it is an Euler-Beta function
\begin{equation}
\begin{aligned}
\left.J_{\al, \be}^n\right|_{s_\para^2 = 0} &= \frac{B_{\al, \be}}{\left|y_\perp\right|^{2(\al + \be) - n}}{}_2F_1\left(\al, \al + \be - \frac{n}{2}; \al + \be; 1 - \frac{x_\perp^2}{y_\perp^2}\right) \ , \\
\left.J_{\al, \be}^n\right|_{y_\perp^2 = 0} &= \frac{B_{\be, n/2 - \be}}{\left(s^2 + x_\perp^2\right)^{\al + \be - n/2}}{}_2F_1\left(\frac{n}{2} - \be, \al + \be - \frac{n}{2}; \frac{n}{2}; \frac{s^2}{s^2 + x_\perp^2}\right) \ , \\
\left.J_{\al, \be}^n\right|_{x_\para^2 = y_\para^2 = 0} &= \frac{B_{n/2 - \al, n/2 - \be}}{\left|s\right|^{2(\al + \be) - n}} \ .
\end{aligned}
\end{equation}

With these integrals we can proceed to find correlators in $3 - \e$ dimensions. The hypergeometric functions can be expanded using the \textbf{HypExp} Mathematica package \cite{0507094}. As argued previously, bulk-bulk correlators are finite, so for these correlator we consider exactly three dimensions. For our purposes we need
\begin{equation*} \hspace{-10px}
\begin{aligned}
J_{1, 1}^2 &= \frac{2}{\sqrt{ \left( s^2 + x_\perp^2 + y_\perp^2 \right)^2 - 4x_\perp^2y_\perp^2 }}\log\left(\frac{s^2 + x_\perp^2 + y_\perp^2 + \sqrt{ \left( s^2 + x_\perp^2 + y_\perp^2 \right)^2 - 4x_\perp^2y_\perp^2 }}{2x_\perp y_\perp}\right) \ , \\
J_{3/2, 1/2}^2 &= \pi\frac{\sqrt{ \left( s^2 + x_\perp^2 + y_\perp^2 \right)^2 - 4x_\perp^2y_\perp^2 } + s^2 + \left( x_\perp - y_\perp \right)^2}{x_\perp \sqrt{2 \left[ \left( s^2 + x_\perp^2 + y_\perp^2 \right)^2 - 4x_\perp^2y_\perp^2 \right] \left(\sqrt{ \left( s^2 + x_\perp^2 + y_\perp^2 \right)^2 - 4x_\perp^2y_\perp^2 } +s^2+x_\perp^2+y_\perp^2\right)}} \ .
\end{aligned}
\end{equation*}

\section{Stress-energy tensor correlator at order $\hg$} \label{App:SE}

The loop integral we wish to find can be written in terms of the master integral (\ref{eq:LoopInt})
\begin{equation} 
\begin{aligned}
\langle\ph^4(x)T^{a\perp}(y)\rangle_1 &= -\frac{\hat{g}}{4!}16N(N + 2) \left\{ (N + 2)\langle\ph^2(y)\rangle^0 K^a + \rig \\
\eq \lef + 2\left(2A_d\right)^4 \left( \p_{y_\perp}\langle\ph(x)\ph(y)\rangle^0\p^a_{y_\para} + \p^a_{y_\para}\langle\ph(x)\ph(y)\rangle^0\p_{y_\perp} \right) I_{3\D, \D}^{d - 1} + \rig \\
\eq \lef - \left(2A_d\right)^4\xi\p^a_{y_\para}\p_{y_\perp} \left[ (N + 2)\langle\ph^2(x)\rangle I_{2\D, 2\D}^{d - 1} + 4\langle\ph(x)\ph(y)\rangle I_{3\D, \D}^{d - 1} \right] \right\} \ ,
\end{aligned}
\end{equation}

where we factored out factors of $2A_d$ from the correlators, and where $K^a$ is the loop integral
\begin{equation*}
\begin{aligned}
K^a &\equiv \int d^{d - 1}z_\para \left( \langle\ph(x)\hp(z)\rangle^0 \right)^2 \p^a_{y_\para}\langle\hp(z)\ph(y)\rangle^0\p_{y_\perp}\langle\hp(z)\ph(y)\rangle^0 = -4\left(2A_d\right)^4\D^2y_\perp\left(L_{2\D, 2(\D + 1)}^{d - 1}\right)^a \ ,
\end{aligned}
\end{equation*}

where $L_{\al, \be}^{n}$ is the integral\footnote{Here we shifted the integration variable as $z_\para^a \rightarrow z_\para^a + y_\para^a$.}
\begin{equation}
\begin{aligned}
\left(L_{\al,\be}^n\right)^a &\equiv \int_{\mathbb{R}^n}d^nz_\para\frac{z_\para^a}{\left[\left(s_\para - z_\para\right)^2 + x_\perp^2\right]^\al\left(z_\para^2 + y_\perp^2\right)^\be} \ .
\end{aligned}
\end{equation}

This integral is solved in a similar manner as $I_{\al,\be}^n$. After a Feynman parametrization followed by the shift $z_\para^a \rightarrow z_\para^a + us_\para^a$
\begin{equation*}
\begin{aligned}
\left(L_{\al,\be}^n\right)^a &= \frac{\G_{\al + \be}}{\G_\al\G_\be}\int_{0}^1duu^{\al - 1}(1 - u)^{\be - 1}\int_{\mathbb{R}^n}d^nz_\para\frac{z_\para^a + us_\para^a}{\left(z_\para^2 + u(1 - u)s_\para^2 + ux_\perp^2 + (1 - u)y\perp^2\right)^{\al + \be}} \ .
\end{aligned}
\end{equation*}

Please note that the denominator is an even function. This means that the term with $z_\para^a$ is an uneven function. Since we are integrating over an even interval it is thus zero. The other term can be integrated over $z_\para$ using (\ref{eq:MyFavInt})
\begin{equation}
\begin{aligned}
\left(L_{\al,\be}^n\right)^a &= s_\para^a\eta^n_{\al\be}J^{n + 2}_{\al + 1, \be} \ ,
\end{aligned}
\end{equation}

Here $\eta^n_{\al\be}$ and $J^{n}_{\al, \be}$ are the same as in (\ref{eq:LoopInt}). We have now computed all of the loop integrals. What is left is the derivatives. Perpendicular derivatives can be performed after integration, but the parallel ones needs to be done prior to integration if we wish to consider small parallel distance. Acting with a parallel derivative on $J^n_{\al\be}$ shifts its arguments
\begin{equation}
\begin{aligned}
\p^a_{y_\para}J^{n}_{\al\be} &= \left( 2(\al + \be) - n \right) s_\para^aJ^{n + 2}_{\al + 1, \be + 1} \ .
\end{aligned}
\end{equation}

For simplicity, let us label operators with their position, and scalar correlators with their corresponding scaling dimension (as well as factor out an $A_d$ from their numerators)
\begin{equation}
\begin{aligned}
\langle\ph^2_x\rangle_\D &\equiv \frac{1}{\left|2x_\perp\right|^{2\D}} \ , \\
\langle\ph_x\ph_y\rangle_\D &\equiv \frac{1}{\left|x - y\right|^{2\D}} + \frac{1}{\left|\tilde{x} - y\right|^{2\D}} \ ,
\end{aligned}
\end{equation}

where $\langle\ph_x\ph_y\rangle_\D$ satisfy
\begin{equation} \label{eq:DerCorr}
\begin{aligned}
\p_{y_\para}^a\langle\ph_x\ph_y\rangle_\D &= 2\D s_\para^a\langle\ph_x\ph_y\rangle_{\D + 1} \ .
\end{aligned}
\end{equation}

In these notations the $\ph^4 - T^{a\perp}$ correlator is
\begin{equation*} \hspace{-10px}
\begin{aligned}
\langle\ph^4(x)T^{a\perp}(y)\rangle_1 &= -\frac{N(N + 2)\left(2A_d\right)^5\hat{g}}{3}s_\para^a \left[ (N + 2)G_1 + 2 \left( G_2 + G_3 \right) - (N + 2)\x G_4 - 4\x \left( G_5 + G_6 \right) \right] \ ,
\end{aligned}
\end{equation*}
\begin{equation}
\begin{aligned}
G_1 &= -4\D^2y_\perp\langle\ph^2_x\rangle_\D\eta^{d - 1}_{2\D, 2(\D + 1)}J^{d + 1}_{2\D + 1, 2(\D + 1)} \ , \\
G_2 &= (8\D - d + 1)\p_{y_\perp}\langle\ph_x\ph_y\rangle_\D\eta^{d - 1}_{3\D, \D}J^{d + 1}_{3\D + 1, \D + 1} \ , \\
G_3 &= 2\D\langle\ph_x\ph_y\rangle_{\D + 1}\eta^{d - 1}_{3\D, \D}\p_{y_\perp}J^{d - 1}_{3\D, \D} \ , \\
G_4 &= (8\D - d + 1)\langle\ph^2_x\rangle_\D\eta^{d - 1}_{2\D, 2\D}\p_{y_\perp}J^{d + 1}_{2\D + 1, 2\D + 1} \ , \\
G_5 &= 2\D\eta^{d - 1}_{3\D, \D}\p_{y_\perp} \left( \langle\ph_x\ph_y\rangle_{\D + 1}J^{d - 1}_{3\D, \D} \right) \ , \\
G_6 &= (8\D - d + 1)\eta^{d - 1}_{3\D, \D}\p_{y_\perp} \left( \langle\ph_x\ph_y\rangle_{\D}J^{d + 1}_{3\D + 1, \D + 1} \right) \ .
\end{aligned}
\end{equation}

This we will first expand in $\e$ (due to the factor of $y_\perp$ in front of the loop integral in $G_1$), and then consider the boundary limit, i.e. we use a cutoff, to reproduce \eqref{eq:BoundLim}.

\bibliographystyle{utphys}
\footnotesize
\bibliography{paper}	
	
\end{document}